\documentclass[reprint,superscriptaddress,eqseqnum,author-year]{revtex4-1}
\usepackage{amsmath,amsthm,amsfonts,amssymb}
\usepackage{graphicx}
\usepackage{color}
\usepackage{natbib}
\usepackage{mciteplus}

\newcommand{\lambdabold}{\mbox{\boldmath{$\lambda$}}}
\newcommand{\ellbold}{\mbox{\boldmath{$\ell$}}}

\newcommand{\sigmabold}{\mbox{\boldmath{$\sigma$}}}
\newcommand{\Sigmabold}{\mbox{\boldmath{$\Sigma$}}}
\renewcommand{\vec}[1]{{\mathbf #1}}        

\newcommand{\ten}[1]{{\mathbf{#1}}}
\newcommand{\half}{\textstyle{\frac{1}{2}}}

\newcommand{\Id}{\ten{I}}

\begin{document}
 
\title{Transient shear banding in the nematic dumbbell model of liquid
  crystalline polymers}

\author{J.~M.~Adams} \affiliation{Department of Physics, University of
  Surrey, Guildford, GU2 7HX, United Kingdom.}

\author{D.~Corbett}
\affiliation{School of Chemical Engineering and Analytical Science, The University of Manchester, Oxford Road, Manchester M13 9PL, United Kingdom.}

\date{\today}

\begin{abstract}
  In the shear flow of liquid crystalline polymers (LCPs) the nematic
  director orientation can align with the flow direction for some
  materials, but continuously tumble in others. The nematic dumbbell
  (ND) model was originally developed to describe the rheology of
  \emph{flow-aligning} semi-flexible LCPs, and flow-aligning LCPs are
  the focus in this paper. In the shear flow of monodomain LCPs it is
  usually assumed that the spatial distribution of the velocity is
  uniform. This is in contrast to polymer solutions, where highly
  non-uniform spatial velocity profiles have been observed in
  experiments. We analyse the ND model, with an additional gradient
  term in the constitutive model, using a linear stability
  analysis. We investigate the separate cases of constant applied
  shear stress, and constant applied shear rate. We find that the ND
  model has a transient flow instability to the formation of a
  spatially inhomogeneous flow velocity for certain starting
  orientations of the director. We calculate the spatially resolved
  flow profile in both constant applied stress and constant applied
  shear rate in start up from rest, using a model with one spatial
  dimension to illustrate the flow behaviour of the fluid. For low
  shear rates flow reversal can be seen as the director realigns with
  the flow direction, whereas for high shear rates the director
  reorientation occurs simultaneously across the gap. Experimentally,
  this inhomogeneous flow is predicted to be observed in flow reversal
  experiments in LCPs.
\end{abstract}
\maketitle

\section{Introduction}

Thermotropic liquid crystalline polymers (LCPs) have a variety of
molecular architectures: ranging from rigid rod-like objects, to
slightly bent rods and semiflexible chains
\cite{doi:10.1146/annurev.fluid.34.082401.191847, Maffetone1992,
  greco1997}. LCPs can be processed into strong, stiff, light weight
fibres, and optical devices. Hence their alignment induced by flow has
been widely studied. They also have applications in electro-optic
devices where they allow tuning of the device properties such as
thermal stability, or viscosity of the device \cite{Mayer2002}.

In the nematic phase LCPs are typically classified according to the
response of the preferred orientation of the nematic mesogens (the
\textit{director}) to the shear flow. In flow tumbling systems the
director continuously rotates in response to a shear strain. In flow
aligning systems the director rotates to approach a steady state angle
aligned in the flow direction for prolate polymer conformations. For
example experimental work on monodomains of rod-like LCPs shows that
they typically exhibit director tumbling
\cite{doi:10.1021/ma00009a063}. Semiflexible chains are more likely to
be flow aligning \cite{semenov1987}. Conoscopy studies of monodomains
of flexible LCPs in shear flow has shown them to be flow aligning
\cite{doi:10.1021/ma0018493,ugaz1998}. These studies have not
investigated the spatial velocity profile in the flow gradient
direction of the rheometer. These two states have been modelled using
the Leslie-Ericksen transversely isotropic fluid model
\cite{lesie1968,ericksen1960}.

The rheology of rod-like LCPs has been successfully modelled by Doi
\cite{doi1984}, and polydomain systems by the Larsen-Doi model
\cite{larson1991}. More flexible LCPs have been modelled using a
slightly bending rod model \cite{greco1995} which is capable of
describing the transition between flow aligning and tumbling behaviour
\cite{greco1997}. Theoretical models typically assume that the flow is
spatially homogeneous, i.e. having a uniform shear rate
\cite{ugaz2001, PhysRevE.68.061704}. Textures in the orientation of
the director (e.g. \cite{gleeson1992}), including a banded structure
in the velocity direction have been predicted using models of rod-like
LCPs, and some of these have included spatial variation in the shear
rate \cite{GREEN200934}. However, the corresponding models have not
been developed for flow-aligning semiflexible LCPs. The rheology of
semiflexible chains has been described theoretically
\cite{semenov1987}, such as through a generalized Rouse model
\cite{long2000}, and a generalized nematic dumbbell model
\cite{Maffetone1992} which is where we will focus in this paper.

The formation of a spatially inhomogeneous flow velocity in polymer
solutions in the flow gradient direction during shear flow, called
\textit{shear banding}, had been long predicted in the Doi-Edwards
model due to a non-monotonic constitutive curve \cite{DE1989}. This
had not been found experimentally until recently
\cite{Tapadia2006}. Theoretically it was shown that a non-monotonic
constitutive curve was not necessary for the formation of shear bands
\cite{PhysRevLett.102.067801}, and that the fluid may be transiently
unstable to the formation of shear bands
\cite{:/content/sor/journal/jor2/55/5/10.1122/1.3610169,PhysRevLett.102.067801}
and even fracture \cite{Agimelen2013}. Analysis of the curvature of
the homogeneous stress response with respect to the strain and the
strain rate can predict the shear banding instability for some
constitutive models
\cite{:/content/sor/journal/jor2/58/1/10.1122/1.4842155,PhysRevLett.110.086001}.
LCP models might also be expected to have an inhomogeneous velocity
profile under suitable conditions.

Orientational banding, i.e.  variation in the director orientation in
response to applied shear strain, is common in LCPs. It is observed in
flow reversal experiments \cite{mather2000}. Crosslinked LCPs that
form a continuous network are called liquid crystal elastomers
(LCEs). LCEs exhibit orientational bands in the director, induced by
deformation, in numerous phases including the nematic
\cite{MACP:MACP677} and smectic phase
\cite{RefWorks:doc:598315c1e4b055eb47de58be, PhysRevE.73.031706,
  RefWorks:doc:598316b9e4b055eb47de58e4}. The formation of the
microstructure in response to mechanical deformation is due to their
unusually soft mechanical response. For certain soft deformations they
deform at virtually no energy cost \cite{warner2003}. This soft
elastic behaviour is accompanied by the formation of spatial
microstructure, and can be traced back to the non-convex shape of the
free energy surface. This soft elastic behaviour is present in the
mechanical response of the nematic dumbbell model \cite{C2SM26868J}.

The linear stability analysis used to examine the transient behaviour
in polymer solutions
\cite{:/content/sor/journal/jor2/55/5/10.1122/1.3610169,:/content/sor/journal/jor2/58/1/10.1122/1.4842155}
can be applied to the flow behaviour of the nematic dumbbell model, to
understand their transient flow instability. The nematic dumbbell
model provides a link between shearbanding and the formation of
microstructure in LCPs. It also gives a possible dynamical model of
the formation of microstructure in LCEs.

This paper is organised as follows. The constitutive equations of the
ND model are introduced in \S\ref{sec:ND}, converted into
dimensionless units, and some suitable values of the model parameters
discussed. The response of the ND model to an imposed shear rate is
then calculated in \S\ref{sec:SS}. The transient response of the ND
model is analysed using linear stability analysis in \S\ref{sec:LSA}
and found to be transiently unstable. The resulting spatiallly
resolved velocity profile in start up flow is calculated in
\S\ref{sec:srm} using a 1-D spatially resolved model. The relation of
the ND model to experimental work and related constitutive models is
discussed in \S\ref{sec:disc}.

\section{The Nematic Dumbbell Model}
\label{sec:ND}
Maffettone and Marrucci developed the nematic dumbbell (ND) model to
describe the rheology of flow-aligning semiflexible LCPs
\cite{Maffetone1992}. They derive the constitutive model for the
polymer shape tensor as follows.
\begin{eqnarray}
\frac{d\langle \vec{R}\vec{R}\rangle}{dt} &=& \ten{K}\cdot\langle \vec{R} \vec{R} \rangle + \langle\vec{R}\vec{R} \rangle\cdot \ten{K}^T\nonumber\\&& + \frac{2 N b^2 \Id}{\tau} - \frac{3}{1-S} \times\nonumber\\
\frac{1}{\tau}\Bigg[ 2 \langle \vec{R} \vec{R} \rangle &-& \frac{3 S}{1+2 S}(\vec{n}\vec{n}\cdot \langle \vec{R}\vec{R} \rangle + \langle \vec{R}\vec{R} \rangle \cdot \vec{n}\vec{n} )\Bigg]
\end{eqnarray}
where $\vec{R}$ is the end-to-end span of the polymer,
$\langle \cdot \rangle$ denotes an ensemble average over many polymer
chains in a volume element,
$\ten{K} = \frac{\partial \vec{v}}{\partial \vec{x}}$ is the velocity
gradient tensor, $S$ is a scalar liquid crystal order parameter,
$\tau$ is the polymer relaxation time, $N$ is the number of Kuhn
segments in the polymer, $b$ is the persistence length, and $\vec{n}$
is the liquid crystalline director. It will be assumed here that we
are deep in the nematic phase, so the nematic order $S$ is fixed. The
polymer stress is specified by
\begin{equation}
\sigmabold = \frac{c k_B T}{N b^2} \frac{3}{1-S}\left( \langle \vec{R}\vec{R}\rangle - \frac{3 S}{1+2 S} \vec{n}\vec{n}\cdot \langle \vec{R}\vec{R}\rangle \right)
\end{equation}
where $c$ denotes the number of chains per unit volume, $T$ is the
temperature, and $k_B$ Boltzmann's constant.

In equilibrium the average polymer spans parallel and perpendicular to
the director are given by the following,
\begin{eqnarray}
\langle R^2_\parallel \rangle &=& \ell_\parallel \frac{N b^2}{3}\\
\langle R^2_\perp \rangle  &=& \ell_\perp \frac{N b^2}{3},
\end{eqnarray}
where $\parallel$ denotes the direction parallel to the director,
$\perp$ denotes the direction perpendicular to it, and
$\ell_\parallel = 1+2S$ and $\ell_\perp = 1-S$. In comparing this
model to the literature on liquid crystalline polymers, elastomers and
transient shear banding, it is convenient to adopt a more compact
notation. Using $\ellbold = \Id +(r-1) \vec{n} \vec{n}$,
where $r = \ell_\parallel/\ell_\perp$ and $\Id$ is the identity tensor, we can write the equilibrium
mean square end-to-end vector of a polymer as
\begin{equation}
\langle \vec{R}\vec{R} \rangle = \ellbold \frac{N b^2 \ell_\perp}{3}.
\end{equation}
When a polymer is out of equilibrium we will denote $\langle \vec{R}
\vec{R} \rangle = \ten{W} \frac{\ell_\perp N b^2}{3}$. Using this
notation, and the upper convected Maxwell derivative
\begin{equation}
  \stackrel{\nabla}{\ten{W}} = \frac{d\ten{W}}{dt} - \ten{K}\cdot \ten{W} - \ten{W} \cdot \ten{K}^T
\end{equation}
we can rewrite Maffettone and Marrucci's model as
\begin{eqnarray}
  \stackrel{\nabla}{\ten{W}} &=& \frac{2}{\tau_\perp}\Id-\frac{1}{\tau_\perp}\left( \ten{W} \cdot \ellbold^{-1} + \ellbold^{-1}\cdot \ten{W} \right) + \mathcal{D} \nabla^2 \ten{W}
\label{eqn:NDcc}
\\
\sigmabold&=& G \ellbold^{-1}\cdot \ten{W},
\label{eqn:NDstress}
\end{eqnarray}
where $G=c k_B T$ and $\tau_\perp = \tau \ell_\perp/3$. Maffettone and
Marrucci discuss various circumstances for the response of the
director \cite{Maffetone1992} -- either by using torque balance, or a
strong external field to determine $\vec{n}$. We will focus here on
the case where the director responds very rapidly, so is always an
eigenvector of $\ten{W}$, which ensures that $\sigmabold$ is a
symmetric tensor (so torque balance is satisfied). In principle there
is a separate time scale for the response of the nematic, and the
polymer backbones. However the response of the nematic is so rapid
compared to the polymer that we will assume that it is
instantaneous. Physically the direction of the director is determined
by the torques from the polymer stress, and the fluid viscosity. We
will consider the regime where $G \gg \alpha \tau$, i.e. where the
polymer stress dominates the determination of the director
orientation. Here $\alpha$ is the appropriate viscosity component of
the nematic.

We have included a diffusive term in the constitutive model only
(Eq.~(\ref{eqn:NDcc})). This stress diffusion term is typically
included to remove the history dependence of shear banding
\cite{PhysRevLett.84.642,RADULESCU2000143}. However we note that a
more rigorous approach would include a diffusive term in the force
balance equation \cite{ottingerHansChristian1992Iopd}.

A full description of this system would include the stress
contribution of the high frequency polymer terms
\cite{PhysRevLett.102.067801}, and the nematic mesogens. This would
couple to the director orientation of the liquid crystalline
polymers. To simplify the model here we represent these high frequency
modes as an isotropic Newtonian solvent term. Hence the total stress
is given by
\begin{equation}
\Sigmabold = -p \Id + \sigmabold + 2\eta \ten{D}
\label{eqn:totalstress}
\end{equation}
where $\ten{D} = \half ( \ten{K}+ \ten{K}^T)$, and $\eta$ is the
viscosity for the high frequency modes. This is typical of models used
to investigate shear banding in worm-like micellar systems
\cite{PhysRevE.68.036313}.

\subsection{Dimensionless Units}

We will work in dimensionless units, using $G$ to set the scale for
stress, $\tau_\perp$ to set the time scale, and the rheometer gap $L$
to set the length scale. In these dimensionless units our equations
become 
\begin{eqnarray}
  \stackrel{\nabla}{\ten{W}} &=&2\Id-\left( \ten{W} \cdot \ellbold^{-1} + \ellbold^{-1}\cdot \ten{W} \right) + \mathcal{\tilde{D}} \tilde{\nabla}^2 \ten{W}\\
\tilde{\sigmabold}&=&  \ellbold^{-1}\cdot \ten{W}\\
\tilde{\Sigmabold}&=&-\tilde{p} \Id + \tilde{\sigmabold} + \epsilon\tilde{\ten{D}}
\label{eqn:NDstressdimensionless}
\end{eqnarray}
where $\tilde{\sigmabold}=\sigmabold/G$, $\mathcal{\tilde{D}} =
\mathcal{D}\tau_\perp/L^2$, $\tilde{\nabla} = L \nabla$, $\tilde{t} =
t/\tau_\perp$ and $\ten{\tilde{K}}= \tau_\perp \ten{K}$. The
dimensionless viscosity of the isotropic solvent is $\epsilon =
\frac{\eta}{G \tau_\perp}$. We will drop the $\tilde{\bullet}$ from here on
and work with the dimensionless quantities, including the
dimensionless local shear rate $\tilde{\dot{\gamma}} = \tau_\perp
\dot{\gamma}$.

\subsection{Model Parameters}
\label{sec:modelparams}

To illustrate the behaviour of this model we will need to use
particular viscosities for our calculations. If we take the viscosity
for the LCP to be in the range $1-10\textrm{ Pa s}$
\cite{doi:10.1021/ma00009a063}, and the viscosity of the Newtonian
solvent term to be $\sim 0.1 \textrm{Pa s}$ (e.g. for MBBA
\cite{chandrasekhar1977}), then $\epsilon \sim 0.01$. Since the ND
model is a single mode approximation to the behaviour of a polymer we
expect the qualitative features to be correct, but not the
quantitative details. We will use $r=2$ for the anisotropy of the
LCPs, typical of a side chain polymer. Typical values for the
reptation time for long polymers is $\tau \sim 1$s, and the rheometer
gap is $L\sim 1\textrm{mm}$ \cite{Tapadia2006}.

The magnitude of the diffusion term has been estimated in worm-like
micellar systems \cite{0295-5075-62-2-230}. Here it is found that
$\mathcal{D} \sim 10^{-13}\textrm{m}^2 \textrm{s}^{-1}$, or in
dimensionless units $\hat{\mathcal{D}} \sim 10^{-7}$. It can also be
justified here as a Frank elasticity type term \cite{Adams2008101}. We
will use a artificially larger diffusion constant of
$\hat{\mathcal{D}} \sim 10^{-4}$ as this makes the number of spatial
grid points smaller. However, the phenomenological effects are the
same for smaller diffusion constants.

\section{Simple shear flow}
\label{sec:SS}
We are interested in the creeping flow limit here, where the Reynolds
number is small. From the parameters given in \S\ref{sec:modelparams}
we estimate $\textrm{Re} \approx \rho v L/\eta \approx 0.01$. In this
case the equation of motion reduces to
\begin{equation}
\nabla \cdot \ten{\Sigma} = 0.
\label{eqn:eom}
\end{equation}
The isotropic pressure can be determined from the incompressibility
condition $\nabla \cdot \vec{v}=0$, where $\vec{v}$ is the velocity
field. 

To analyse the behaviour of the ND model we consider its response in a
simple shear flow geometry. We will assume that the fluid is held
between parallel plates at $y=0$ and $y=1$. The fluid velocity will be
of the form $\vec{v} = v(y,t) \vec{x}$, and the local shear rate
\begin{equation}
\dot{\gamma}(y,t) = \partial_y v(y,t).
\end{equation}
Using Eq.~(\ref{eqn:totalstress}) and Eq.~(\ref{eqn:eom}) we find that
\begin{equation}
\Sigma_{xy}(t) = \sigma_{xy} + \epsilon \dot{\gamma}
\label{eqn:stressint}
\end{equation}
where $\Sigma_{xy}(t)$ is the total shear stress, and is independent
of spatial coordinates. We will use Eq.~(\ref{eqn:stressint}) in the
fixed shear stress case later to substitute for the local shear rate.

As a result of the shear flow geometry the stress component
$\Sigma_{zz}$ decouples from the other components, so we will ignore
it here. We will also assume that the director remains in the $xy$
plane.  Assuming that $\vec{n}$ is the eigenvector of $\ten{W}$ with
the largest eigenvalue $\lambda$ (for mechanical stability when
$r>1$), then the remaining equations can be written as
\begin{eqnarray}
\label{eqn:ceqnxx}
\dot{W}_{xx}- \mathcal{D} \partial^2_yW_{xx} &=& 2 W_{xy} \dot{\gamma} + 2 \left( 1 - \left[\ellbold^{-1}\cdot \ten{W}\right]_{xx}\right)  \\
\label{eqn:ceqnyy}
\dot{W}_{yy}- \mathcal{D} \partial^2_yW_{yy}&=&2 \left( 1 - \left[\ellbold^{-1}\cdot \ten{W}\right]_{yy}\right)  \\
\label{eqn:ceqnxy}
\dot{W}_{xy}- \mathcal{D} \partial^2_yW_{xy} &=& W_{yy} \dot{\gamma} - 2\left[\ellbold^{-1}\cdot \ten{W}\right]_{xy}
\end{eqnarray}
where
$\ellbold^{-1}\cdot \ten{W} = \ten{W} + \left(\frac{1}{r}-1
\right)\lambda \vec{n}\vec{n}$.
The components of this dot product give rise to the non-linear
behaviour of this model.

\subsection{Eigenbasis equations}

Calculating the properties of the ND model in the steady state, and
for a homogeneous system ($\mathcal{D}=0$) is simplified if we work in
the basis of the director, $\vec{n}$. In two dimensions, we can write
the director and its perpendicular component as
\begin{eqnarray}
\vec{n} &=& ( \cos \theta, \sin \theta)\\
\vec{n}_\perp &=& (-\sin \theta, \cos \theta)\\
\Rightarrow \ten{W} &=& W_1 \vec{n} \vec{n} + W_2 \vec{n}_\perp \vec{n}_\perp.
\end{eqnarray}
The equations for $W_1, W_2$ and $\theta$ can be found from
Eq.~(\ref{eqn:NDstressdimensionless}) by resolving along
$\vec{n}\vec{n}$, $\vec{n}_\perp\vec{n}_\perp$ and
$\vec{n}\vec{n}_\perp$. The constitutive equations become
\begin{eqnarray}
\dot{W_1} &=& 2 - \frac{2 W_1}{r} + W_1 \dot{\gamma} \sin 2 \theta \label{eqn:AW1}\\
\dot{W_2} &=&2 (1 - W_2 - W_2 \dot{\gamma} \cos \theta \sin\theta)\label{eqn:AW2}\\
\dot{\theta} &=&\frac{\dot{\gamma}((W_2-W_1)+(W_2+W_1)\cos 2\theta)}{2(W_1-W_2)}.\label{eqn:Atheta}
\end{eqnarray}
The components of $\ten{W}$ can be interpreted as the extension of the
conformation tensor along the director, $W_1$ and perpendicular to the
director $W_2$. Note that since the director is a quadrupolar object,
the angle $\theta$ and $\theta+\pi$ correspond to the same physical
state.

\subsection{Steady state}
 
The steady state behaviour of the homogeneous ND model for imposed
shear rate has been solved in the large shear rate limit
$\dot{\gamma}\rightarrow \infty$ by Maffettone and Marrucci in
\cite{Maffetone1992}. We solve the elastic limit in appendix
\ref{app:elasticlimit}, and discuss the small amplitude and the small
amplitude oscillatory shear response in appendix
\ref{app:saresponse}. In this section we give an exact solution of the
steady state equations for the stress. First we substitute for
$\sigmabold$ from Eq.~(\ref{eqn:NDstress}) into
Eqs. (\ref{eqn:ceqnxx}, \ref{eqn:ceqnyy}, \ref{eqn:ceqnxy}), which in
the steady state with $\mathcal{D} = 0$ gives
\begin{eqnarray}
\sigma_{xx} &=& (1+\dot{\gamma} W_{xy})\\
\sigma_{yy} &=& 1\\
\sigma_{xy} &=& \frac{\dot{\gamma} W_{yy}}{2}.
\end{eqnarray}
Then to determine the three components of $\ten{W}$ we use the trace
and determinant of Eq.~(\ref{eqn:NDstress}), and the fact that
$\sigmabold$ and $\ten{W}$ must commute, i.e.
\begin{eqnarray}
\textrm{tr}(\ten{W}) -r \sigma_1 - \sigma_2&=& 0\\
\textrm{det}(\sigmabold)r-\textrm{det}(\ten{W}) &=& 0\\
\ten{W} \cdot \sigmabold = \sigmabold \cdot \ten{W}
\end{eqnarray}
where $\sigma_1$ and $\sigma_2$ are the eigenvalues of $\sigmabold$. 
Solving these equations for the components of $\ten{W}$ yields
\begin{eqnarray}
W_{xx} &=&\frac{(1+r^2)(2+r \dot{\gamma}^2)}{2(1+r)}\!+\!\frac{(r-1)\sqrt{r}\dot{\gamma}}{2}\sqrt{4+r \dot{\gamma}^2}\\
W_{yy} &=&\frac{2 r}{1+r}\\
W_{xy} &=&\frac{\sqrt{r}}{2}\left(\sqrt{r} \dot{\gamma}+\frac{r-1}{r+1}\sqrt{4+r \dot{\gamma}^2}\right).
\end{eqnarray}
Hence the total shear stress in the steady state is
\begin{equation}
\Sigma_{xy} = \frac{\dot{\gamma}r}{r+1}+\epsilon \dot{\gamma}.
\label{eq:sigma}
\end{equation}
Shear banding in the steady state is predicted in models that have a
non-monotonic constitutive curve, i.e.
$\partial_{\dot{\gamma}} \Sigma_{xy}<0$ \cite{Radulescu1999}. The ND
model has a linear stress-shear rate behaviour there is therefore no expectation
of spatially inhomogeneous flow in the steady state.

The equilibrium value of the director angle with respect to the $x$
axis, $\theta$, can be found from the eigenbasis equations
(\ref{eqn:AW1}),(\ref{eqn:AW2}) and (\ref{eqn:Atheta}). In the steady
state we set $\dot{\theta}=\dot{W}_{1}=\dot{W}_{2}=0$.  Solving
Eqs.~(\ref{eqn:AW1}) and ~(\ref{eqn:AW2}) for $W_{1}$ and $W_{2}$ as
functions of $\dot{\gamma}$ and inserting the result into
Eq.~(\ref{eqn:Atheta}) gives:
\begin{equation}
(r+1)\cos 2\theta = (r-1)+r\dot{\gamma}\sin 2\theta.
\end{equation}
Let $t=\tan\theta$, in terms of which $\cos 2\theta=(1-t^{2})/(1+t^{2})$ and $\sin 2\theta = 2t/(1+t^{2})$, which gives a quadratic for $t$:
\begin{equation}
2rt^{2}+2r\dot{\gamma}t-2=0
\end{equation}
i.e.
\begin{equation}
\tan\theta=-\frac{\dot{\gamma}}{2}\pm\sqrt{\left(\frac{\dot{\gamma}}{2}\right)^{2}+\frac{1}{r}}.
\label{eqn:directorSS}
\end{equation}
A linear stability analysis can be used to determine which of these
solutions is stable under shear flow. Suppose that only $\theta$
varies and $W_1$, $W_2$ remain fixed at their steady state values
(corresponding to rotating the polymer around its steady state, but
not stretching it). In this case the negative solution is only stable
for large values of $\dot{\gamma}>0$ whereas the positive solution is
stable for all values of $\dot{\gamma}>0$. Swapping to
$\dot{\gamma}<0$ results in changing over the stability of the two
solutions. We find the positive root occurs in the steady state in our
numerical calculations.

\section{Linear Stability Analysis}
\label{sec:LSA}

Linear stability analysis (LSA) of the constitutive equations has been
used to determine whether the homogeneous state is unstable to the
formation of spatial structure, in particular shear bands. For
example, this has been done for the Diffusive Johnson-Segalman model
in the \emph{steady state} \cite{WILSON2006181}. LSA of spatial
perturbations around the \emph{time dependent} transient state for
start up flow of the Diffusive Johnson-Segalman and the Diffusive
Rolie-Poly models \cite{LIKHTMAN20031} have been carried out
\cite{PhysRevE.68.036313,
  :/content/sor/journal/jor2/55/5/10.1122/1.3610169}. Moorcroft and
Fielding have developed a criterion to detect transient shear banding
of complex fluid flow based on LSA
\cite{:/content/sor/journal/jor2/58/1/10.1122/1.4842155,
  PhysRevLett.110.086001}. We will use the eigenvalues obtained from a
LSA here, rather than the criterion of Moorcroft and Fielding as some
of the assumptions required in the derivation are not satisfied for
the ND model. In particular the determinant of the stability matrix
changes sign, and the eigenvalues can appear in complex conjugate
pairs. This is discussed in appendix \ref{app:LSAeigenvalues}.

We give a brief summary here of the relevant stability analysis using
the notation of Ref.
\cite{:/content/sor/journal/jor2/58/1/10.1122/1.4842155}. The
constitutive equations (\ref{eqn:ceqnxx}), (\ref{eqn:ceqnyy}), and
(\ref{eqn:ceqnxy}) can be rewritten in terms of
$\vec{s} = (W_{xy}, W_{xx}, W_{yy})$ as
\begin{equation}
\partial_t \vec{s} = \vec{Q}(\vec{s}, \dot{\gamma})+D \partial_y^2 \vec{s}
\label{eqn:ccs}
\end{equation}
where $\vec{Q}$ is the function that specifies the constitutive
model. The total shear stress is given by
\begin{equation}
\Sigma(t) =  f(\vec{s}) + \epsilon \dot{\gamma}
\label{eqn:sigmas}
\end{equation}
where $f(\vec{s})$ is determined by the dot product of $\ten{W}$ and
$\ellbold^{-1}$ by Eq.~(\ref{eqn:NDstress}).  Assuming that
$\vec{s}$ obeys the Neumann boundary condition
$\partial_y \vec{s} = 0$ at $y=0$ and $L$, then the spatial
fluctuations in $\vec{s}$ and $\dot{\gamma}$ about their homogeneous
values can be written as
\begin{eqnarray}
\dot{\gamma}(y,t) = \dot{\gamma}_0(t) + \sum_{n=1}^\infty \delta\dot{\gamma}_n(t) \cos(n \pi y/L)\\
\vec{s}(y,t) = \vec{s}_0(t) + \sum_{n=1}^\infty \delta\vec{s}_n(t) \cos(n \pi y/L)
\end{eqnarray}
where $\delta\vec{s}_n$ and $\delta \dot{\gamma}_n$ are the Fourier
coefficients for the fluctuations, and $\dot{\gamma}_0$ and
$\vec{s}_0$ are the homogeneous base states. We will examine the
stability under two different conditions: step shear stress, in which
the total shear stress $\Sigma$ is held fixed, and step shear rate, in
which the average shear rate $\overline{\dot{\gamma}}$ is held
fixed. The stability of the system to spatial fluctuations can be
obtained from first calculating the base state $\vec{s}_0(t)$ which is
obtained from the zeroth order equations (no fluctuations):
\begin{eqnarray}
\Sigma_0(t) &=& f(\vec{s}_0(t)) + \epsilon \dot{\gamma}_0(t)\\
\dot{\vec{s}}_0 &=& \vec{Q}(\vec{s}_0,\dot{\gamma}_0).
\end{eqnarray}
To find the fluctuations around this base state, $\delta{\vec{s}}_n$
we use the first order equations:
\begin{eqnarray}
0 &=& \vec{p}\cdot \delta\vec{s}_n + \epsilon \delta \dot{\gamma}_n
\label{eqn:dgamfluctuations}\\
\dot{\delta\vec{s}}_n &=&\ten{M}(t) \cdot \delta\vec{s}_n + \vec{q} \delta \dot{\gamma}_n
\label{eqn:dsfluctuations}
\end{eqnarray}
where $\ten{M} = \partial_\vec{s}\vec{Q}$,
$\vec{p} = \partial_\vec{s} f(\vec{s})$ and
$\vec{q} = \partial_{\dot{\gamma}} \vec{Q}$. Combining these two
equations gives
\begin{eqnarray}
  \dot{\delta \vec{s}}_n =  \ten{P}\cdot \delta \vec{s}_n.
\label{eqn:Mfluctuations}
\end{eqnarray}
where 
\begin{equation}
  \ten{P}(t) = \left( \ten{M}(t) -\frac{1}{\epsilon} \vec{q} \vec{p} \right).
\label{eqn:stabmat}
\end{equation}
The eigenvalues of the matrix $\ten{P}$ determine whether fluctuations
grow or shrink. If the real part of an eigenvalue of $\ten{P}$ is
positive then the fluctuations along the corresponding eigenvector
will grow with time. Conversely if they have negative real part then
the fluctuations will decay with time. We will denote real part of the
eigenvalue with largest real part as $\omega$.

\subsection{Step shear stress}

\begin{figure*}[!htbp]
\begin{center}
\includegraphics[width = 0.9\textwidth]{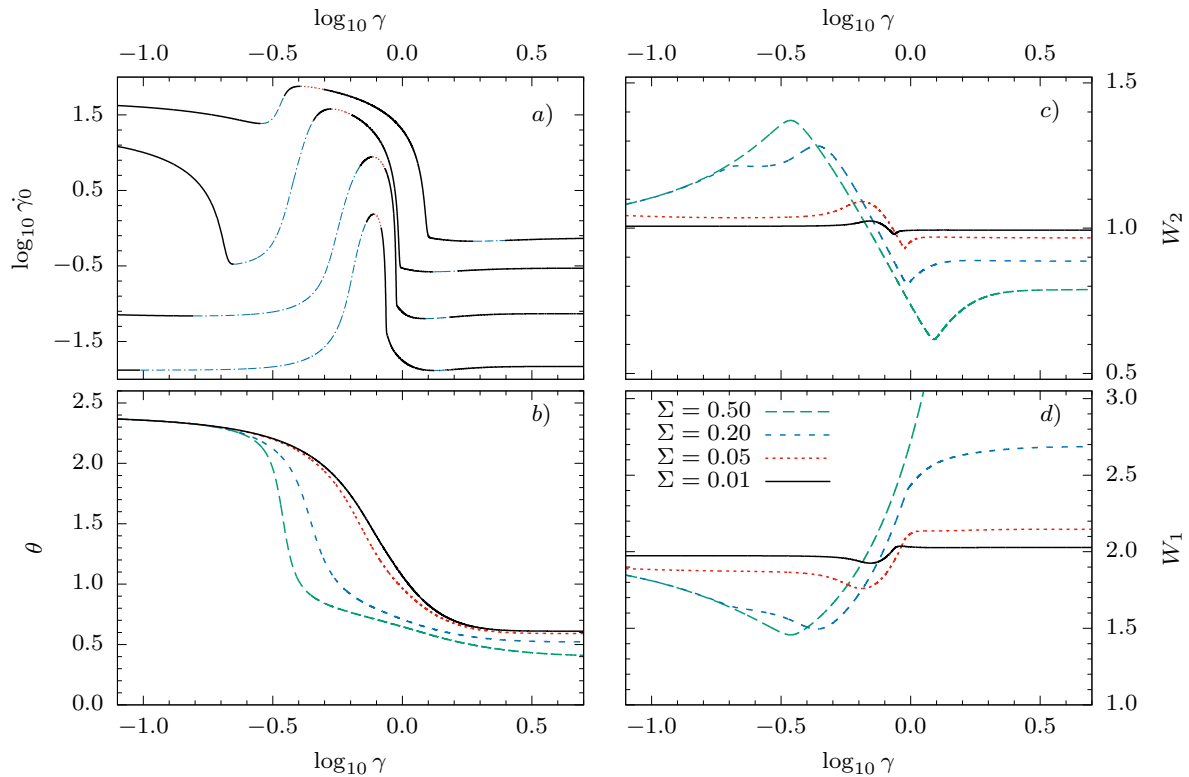}
\end{center}
\caption{The evolution of the ND model assuming homogeneous flow, for
  fixed applied stress, $\theta_0 =2.4$, $\epsilon = 0.01$ and
  $r=2$. (a) shows the strain rate evolution. The solid black lines
  are stable flow, and the dashed regions are unstable. The blue long
  dashing has $d\dot{\gamma}/d\gamma>0$, and the red dashed region has
  $d\dot{\gamma}/d\gamma<0$ (Eq.~(\ref{eqn:upcurve}) and
  (\ref{eqn:dncurve})) . (b) shows the evolution of the director
  angle, (c) and (d) show the $W_2$ and $W_1$ components of the
  polymer shape tensor. Note that (d) shows that $W_1$ shrinks before
  the director rotation, corresponding to compressing the polymers
  along their long axis. After rotation the polymers are then extended
  by the shear flow.}
\label{fig:fixed_stress}
\end{figure*}

The fluid starts in an equilibrium state at $t=0$, and is subjected to
a step $xy$ shear stress of magnitude $\Sigma_0$. The homogeneous
shear rate that arises in response to this stress,
$\dot{\gamma}_0(t)$, can be calculated by numerical solution of the
ordinary differential equations (\ref{eqn:ceqnxx}, \ref{eqn:ceqnyy},
\ref{eqn:ceqnxy}) (setting $\mathcal{D} = 0$) and substituting for
$\dot{\gamma}$ using
\begin{equation}
\dot{\gamma} = \frac{(\Sigma - \sigma_{xy})}{\epsilon},
\label{eqn:stress_strain}
\end{equation}
where $\sigma_{xy}$ can be found in terms of $\ten{W}$ from
Eq.(\ref{eqn:NDstress}). LSA gives us the condition for the
development of spatial fluctuations. The fluctuations around the base
state obey Eq.~(\ref{eqn:Mfluctuations}). These fluctuations obey the
same dynamical equation as the base state $\vec{s}_0$, so it can be
shown that the condition for the growth of fluctuations is
\cite{:/content/sor/journal/jor2/55/5/10.1122/1.3610169}
\begin{equation}
\frac{d^2 \dot{\gamma}_0}{dt^2}/\frac{d \dot{\gamma}_0}{dt}>0,
\end{equation}
i.e. we are looking for both upward sloping and upward curving shear
rate, or downward sloping and downward curving shear rate. The
numerical results of this calculation can be most easily understood by
plotting the shear rate as a function of strain, since
$\dot{\gamma} = \dot{\gamma}(\gamma)$, for different total stress
values. This condition can be converted to strain to give
\begin{equation}
\frac{d\dot{\gamma}}{d\gamma}>0 \;\; \textrm{and} \;\; \frac{d^2 \dot{\gamma}}{d\gamma^2}>-\frac{1}{\dot{\gamma}}\left(\frac{d \dot{\gamma}}{d \gamma}\right)^2
\label{eqn:upcurve}
\end{equation}
or 
\begin{equation}
\frac{d\dot{\gamma}}{d\gamma}<0 \;\; \textrm{and} \;\; \frac{d^2 \dot{\gamma}}{d\gamma^2}<-\frac{1}{\dot{\gamma}}\left(\frac{d \dot{\gamma}}{d \gamma}\right)^2
\label{eqn:dncurve}
\end{equation}

The negative sloping and negative curvature condition is observed in
the ND model (Moorcroft and Fielding comment that it is not observed
in Giesekus or the Rolie-Poly model
\cite{:/content/sor/journal/jor2/58/1/10.1122/1.4842155}). Note that
the condition in strain variables here requires that the curvature
with respect to strain be more negative for more steeply sloped curves
as compared to the corresponding situation with positive
curvature. This is evident in the following numerical calculations.

The constitutive equations in the eigenbasis for the ND model were
solved using the NAG C library \texttt{d02ejc} \cite{NAGlibrary}. This
is an implementation of variable-step backward differentiation
formulae for stiff ordinary differential equations. The stability of
the system is sensitive to the initial orientation of the director
$\theta_0$. For prolate polymer conformation ($r>1$) the director
rotates towards the stable solution of Eq.~(\ref{eqn:directorSS}). For
director angles close to the stable solution there is no flow
instability predicted by LSA. However, if the director angle is close
to the unstable solution of Eq.~(\ref{eqn:directorSS}) then there is a
sharp peak in $\dot{\gamma}_0$. Fig.~\ref{fig:fixed_stress}a) shows
the shear rate as a function of strain for a variety of different
total shear stresses, with a fixed starting angle of
$\theta_0=2.4$. The unstable regions of this curve are highlighted
with a dashed line. Note that there are small regions of
\emph{negative curvature} that are unstable for the ND model. However
the instability arising from the preceding upward sloping and upward
curving region of the shear rate would result in an inhomogeneous
velocity profile, and make the underlying assumption of a spatially
homogeneous state for subsequent regions of the curve invalid.

The peak in the strain can be understood from
Eq.~(\ref{eqn:stress_strain}). As a result of the flow there is a
component of the flow field that gradually rotates the
director. However, due to the alignment of the director the
corresponding polymer shear stress component $\sigma_{xy}$ gradually
falls to zero as the director rotates, and so to maintain the fixed
stress condition the shear rate $\dot{\gamma}$ increases. The peak in
the shear rate occurs when $\sigma_{xy}=0$, where
$\dot{\gamma}=\Sigma/\epsilon$. This expression corresponds to the
peaks in strain rate in Fig.~\ref{fig:fixed_stress}a). The associated
realignment of the director is shown in
Fig.~\ref{fig:fixed_stress}b). The rapid reorientation of the director
results in a stable angle of the director from
Eq.~(\ref{eqn:directorSS}), and resolves the unstable flow.

\begin{figure}[!htbp]
\begin{center}
\includegraphics[width = 0.5\textwidth]{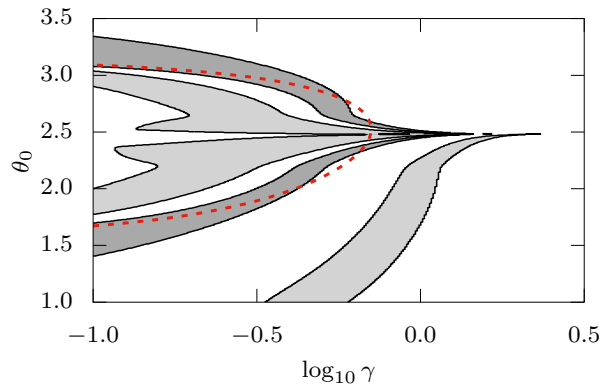}
\end{center}
\caption{The stability of the homogeneous state at fixed stress with
  $\Sigma=0.1$, $r=2$ $\epsilon = 0.01$ as a function of initial angle
  of the director $\theta_0$ and shear strain $\gamma$. The shaded
  area shows $\dddot{\gamma}/\ddot{\gamma}>0$, with light grey for
  $(\dot{\gamma},\dddot{\gamma})>0$ and dark grey for $(\dot{\gamma},\dddot{\gamma})<0$. The dashed
  (red) line shows the maximum strain of the soft mode of an
  LCE in Eq.~(\ref{eqn:trstrain}). }
\label{fig:fixed_stress_sweep}
\end{figure}

The shear flow distorts the equilibrium polymer shape as the flow
progresses. Initially the average conformation of the LCPs are prolate
spheroids with their long axis parallel to the director. However for
the LCPs in Fig.~\ref{fig:fixed_stress} they are compressed along
$\vec{n}$ (i.e $W_1$) and elongated in the perpendicular direction
(i.e. $W_2$), storing elastic energy, before reorientation
(Fig.~\ref{fig:fixed_stress}c) and d)). The rotation of the director
then allows the polymers to release this elastic energy, and the flow
field continues to stretch the polymers along the director.

The instability is sensitive to the initial orientation of the
director. Fig.~\ref{fig:fixed_stress_sweep} shows the region of
instability as a function of initial angle $\theta_0$ and $\gamma$ for
$\Sigma=0.1$. The correspondence to Fig.~\ref{fig:fixed_stress} can be
seen with the two bands for small strains corresponding to the leading
and trailing edges of the peak in shear rate. The instability is
strongest when the initial director angle is pointed away from the
flow direction. Note that the initial angle $\theta_0$ where there is
a cusp as a function of strain corresponds to $\dot{\theta}=0$ in the
constitutive equations (Eq.~(\ref{eqn:AW1}, \ref{eqn:AW2},
\ref{eqn:Atheta})).

\subsubsection{Relation to soft elasticity}

The shape of the shaded unstable regions in
Fig.~\ref{fig:fixed_stress_sweep} can be understood by comparing them
with the equilibrium model of liquid crystalline elastomers (LCEs),
which is obtained in the elastic limit of the ND model. In this case
an analytical expression for the expected value of this strain of the
soft mode can be calculated from the trace formula used to describe
LCEs. The free energy, $F$, here is given by
\begin{equation}
  F = \half \mu \textrm{Tr} \left[ \lambdabold \cdot \ellbold_0 \cdot \lambdabold \cdot \ellbold^{-1} \right]
\end{equation}
where $\mu$ is the shear modulus, $\lambdabold$ is the deformation
matrix, $\ellbold_0 = \Id + (r-1) \vec{n}_0 \vec{n}_0$ is the
initial polymer shape tensor, and
$\ellbold = \Id +\left(\frac{1}{r} - 1\right) \vec{n}\vec{n}$ is
the current polymer shape tensor \cite{warner2003}. We set
$\vec{n}_0 = (\cos \theta_0, \sin\theta_0,0)$,
$\vec{n} = (\cos \theta, \sin\theta,0)$ and
$\lambdabold = \Id + \hat{\vec{x}}\hat{\vec{y}} \gamma_0$. The free energy $F$
is then minimised with respect to $\theta$ for a fixed strain
$\gamma_0$ and initial angle $\theta_0$. It can be shown that this
expression has minimum in $F$ for $\gamma_0 = 0$ and
\begin{equation}
\gamma_0 = \frac{2(r-1) \sin 2 \theta_0}{(r-1) \cos 2 \theta_0 -(r+1)}.
\label{eqn:trstrain}
\end{equation}
For the initial conditions in Fig.~\ref{fig:fixed_stress}, this
expression gives a value of $\textrm{log}_{10}\gamma_0 \approx -0.15$
which coincides with the peak in the shear rate in
Fig.~\ref{fig:fixed_stress}.

Eq.~(\ref{eqn:trstrain}) predicts that the position of the peak in the
strain rate depends on the initial angle $\theta_0$. A contour of the
strain as a function of the initial angle, $\theta_0$ is shown in
Fig.~\ref{fig:fixed_stress_sweep}. The maximum amplitude of $\gamma_0$
corresponds to the cusp shown in this figure.

\subsection{Step shear rate}

\begin{figure*}[!htb]
\begin{center}
\includegraphics[width = \textwidth]{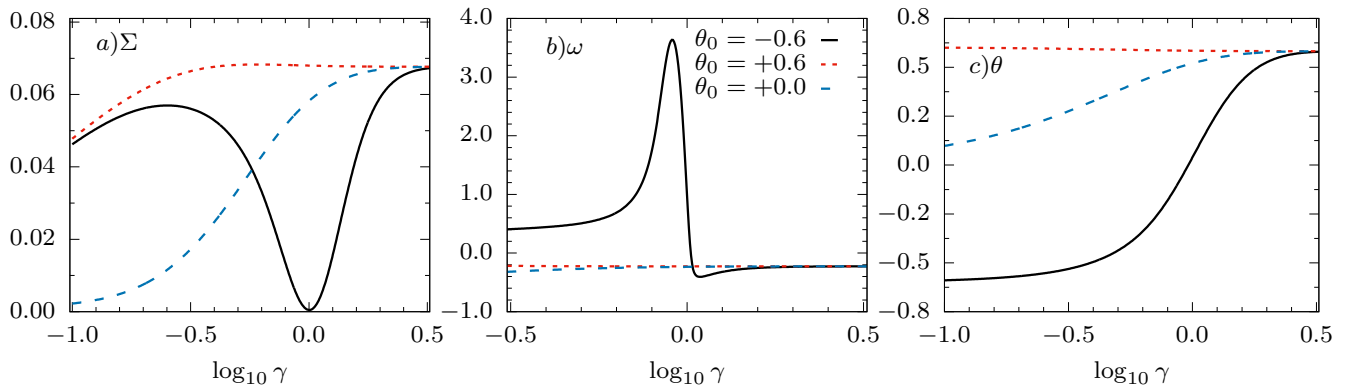}
\end{center}
\caption{The evolution of a) the total stress, b) maximum eigenvalue
  $\omega$, and c) director angle for an imposed shear rate
  $\dot{\gamma}_0 = 0.1$, polymer anisotropy $r=2$ and initial
  director orientations of $\theta_0=-0.6,0.0,0.6$ (black solid, red
  dotted, and blue dashed lines respectively).}
\label{fig:fixedshearrate}
\end{figure*}

We now consider a step shear rate experiment. The fluid starts in its
equilibrium state at $t=0$ and is then subjected to a shear rate
$\dot{\gamma}$ for $t>0$. The stability of the homogeneous base state
to spatially inhomogeneous flow can be found by analysing the
eigenvalues of the matrix $\ten{P}$ given in
Eq.~(\ref{eqn:stabmat}). The behaviour of the fluid for starting
angles of $\theta_0 = 0.6, 0.0$ and $-0.6$ are shown in
Fig.~\ref{fig:fixedshearrate}. The total shear stress is monotonically
increasing for $\theta_0 = 0$ or $0.6$, and $\omega$ remains negative
for all values of shear strain. No radical change of the director
orientation is required here. However, for $\theta_0 = -0.6\equiv \pi-0.6$ the
director undergoes a large rotation towards the flow direction (solid
black line in Fig.~\ref{fig:fixedshearrate} c)). During the rotation
there is a drop in the shear stress, and a simultaneous spike in the
value of $\omega$, a sign of a spatial instability. This indicates
that small perturbations of polymer stress components around the
homogeneous base state should grow here. One difficulty with this
eigenvalue analysis is that we do not know for how long or how
positive the eigenvalues must be in order to cause a spatial
instability. Previous analysis has looked at the integrated area of
the positive region of $\omega$
\cite{:/content/sor/journal/jor2/55/5/10.1122/1.3610169}, however this
is not particularly instructive. For larger values of shear rate the
total stress dips to negative values for the homogeneous state. This
is typical of the behaviour of LCEs during their deformation.

An alternative method of determining the stability of the fluid to
fluctuations for imposed shear rate is presented in appendix
\ref{app:gdotfluct}. The properties of the eigenvalues of this system
of equations make it difficult to use the stability criterion of
Moorcroft et
al. \cite{:/content/sor/journal/jor2/58/1/10.1122/1.4842155}. These
properties are discussed in appendix \ref{app:LSAeigenvalues}.

\section{Spatially resolved model}
\label{sec:srm}

To understand the nature of the instabilities predicted from LSA we
will solve the constitutive equations in Eq.~(\ref{eqn:ceqnxx}),
(\ref{eqn:ceqnyy}) and (\ref{eqn:ceqnxy}) for the 1D case of a planar
shear between two infinite plates at $y=0$ and $y=1$. We will use
Neumann boundary conditions at $y=0,1$
$\frac{\partial W_{\alpha\beta}}{\partial y}=0\,\forall\,\alpha,\beta$
for $\ten{W}$, while we will assume no wall slip and no penetration of
the particles through the wall for the velocity i.e.
$\vec{v} = v(y,t) \vec{x}$. The effect of changing the boundary
conditions in shear banding systems has been explored elsewhere
\cite{Adams2008101}.

In the creeping flow approximation we ignore inertia, so force balance
reduces to Eq.~(\ref{eqn:eom}). Since we only have spatial variation
in the $y$-direction
(i.e. \mbox{$\nabla\equiv\hat{\vec{y}}\frac{\partial}{\partial y}$})
then integrating Eq.~(\ref{eqn:eom}) with respect to $y$ gives
$\Sigma_{xy}(y,t)=\sigma_{xy}+\epsilon\dot{\gamma}=f(t)$, i.e. the
total shear stress is the same at all points across the gap, though it
can vary with time. We will use this condition in the fixed average
shear rate case to calculate the local shear rate as follows
\begin{equation}
\Sigma_{xy}(t) = \sigma_{xy} + \epsilon \dot{\gamma} = \overline{\sigma_{xy}} + \epsilon \overline{\dot{\gamma}},
\label{eqn:localstraingdot}
\end{equation}
where the bar denotes the spatial average
\begin{equation}
\overline{\dot{\gamma}} = \int_0^1 \dot{\gamma}(y,t) dy.
\end{equation}
For a fixed total shear stress $\Sigma_{xy}$ the local shear rate is
given by:
\begin{equation}
  \dot{\gamma}(y,t)=
  (\Sigma_{xy}-\sigma_{xy}(y,t))/\epsilon.
\label{eqn:localstrainstress}
\end{equation}

The inhomogeneity that arises in the flow field can be quantified in
many different ways, such as the difference between the maximum and
minimum shear rates:
$\dot{\gamma}_\textrm{max} - \dot{\gamma}_\textrm{min}$
\cite{:/content/sor/journal/jor2/55/5/10.1122/1.3610169}. We use here
a more robust measure of the inhomogeneity that does not depend so
critically on just two values of the shear rate:
\begin{equation}
\Delta \dot{\gamma} = \int_0^1\left|\dot{\gamma}(y)-\overline{\dot{\gamma}}\right|dy.
\end{equation}
For a system with a uniform shear rate this will be zero, and it will
be positive for non-uniform shear rate profiles.

\subsection{Numerical scheme}

For numerical solution of Eq.~(\ref{eqn:ceqnxx}), (\ref{eqn:ceqnyy})
and (\ref{eqn:ceqnxy}) we use a finite difference scheme with two
staggered uniform grids each with spacing $\Delta y$,
$y_n = y_0 + n \Delta y$. We use the full points
$y_{0},y_{1}\ldots y_{N}$ for the velocity field $v_{x}(y,t)$ and the
half-points $y_{1/2},y_{3/2}\ldots y_{N-1/2}$ for $\ten{W}$,
$\sigmabold$ and $\dot{\gamma}$.

In order to integrate from time $n\Delta t$ to $(n+1)\Delta t$ we
first use the values of $W^{(n)}_{xx},\, W^{(n)}_{xy},\,W^{(n)}_{yy}$
at the current time-step $n\Delta t$ to calculate the values of
$\dot{\gamma}^{(n)}(y_{i/2},n\Delta t)$ with
Eq.~(\ref{eqn:localstraingdot}) for the fixed strain rate, and
Eq.~(\ref{eqn:localstrainstress}) for the fixed stress case. These are
then used in the finite difference form of the constitutive equations
which are integrated forwards in time using the Crank-Nicolson
algorithm~\cite{Press:1993:NRF:563041} to obtain
$W^{(n+1)}_{xx},\, W^{(n+1)}_{xy},\,W^{(n+1)}_{yy}$ at the new
time-step. In addition the values of $\dot{\gamma}^{(n)}(y,t)$ are
integrated spatially to obtain the velocity at each full grid point
$v^{(n)}_{x}(y_{i},n\Delta t)$.

For our chosen value of $\hat{D}=10^{-4}$ we expect a shear band to
have a thickness $l\approx\sqrt{\hat{\cal{D}}}=10^{-2}$. In order to
have roughly $10$ grid points on the interface we should then have
$\Delta y \lesssim 10^{-3}$, i.e. we need $N\gtrsim 10^{3}$ grid
points. We have tested our algorithm for convergence as we change both
$\Delta t$ and $\Delta y$.  To obtain stable and accurate results we
find we need
$\Delta t\approx \Delta y^{2}/(10\hat{\cal{D}})\approx 10^{-3}$.

\subsection{Initial conditions}

The initial conditions have a dramatic effect on the evolution of the
system because they are amplified dramatically as a result of the flow
instability. A small noise term was used to seed the initial
configuration to make the calculations more reproducible. The noise
was set using Fourier harmonics with random amplitudes. High frequency
harmonics result in many interfaces developing, and a more complicated
spatial structure, which eventually becomes uniform as the system
evolves. To keep the spatial structure simple we used the following
initial condition in start up from the relaxed state
\begin{equation}
\ten{W} = {\ellbold}_0 +U_{xy} (\hat{\vec{x}}\hat{\vec{y}} + \hat{\vec{y}}\hat{\vec{x}})
\end{equation}
with the perturbing noise term
\begin{equation}
  U_{xy} =  \xi \cos\frac{\pi y}{L}.
\end{equation}
It was found that a noise amplitude of $\xi=10^{-2}$ was adequate to
trigger the instability reliably. 

Note that the equations solved here are for a parallel plate
rheometer. The curvature of the rheometer has been included elsewhere,
and is found to break the symmetry of the system and determine where
the high and low shear rate bands form \cite{Adams2008101}.

\subsection{Imposed average shear rate}

\begin{figure*}[!htb]
\begin{center}
  \includegraphics[width =\textwidth]{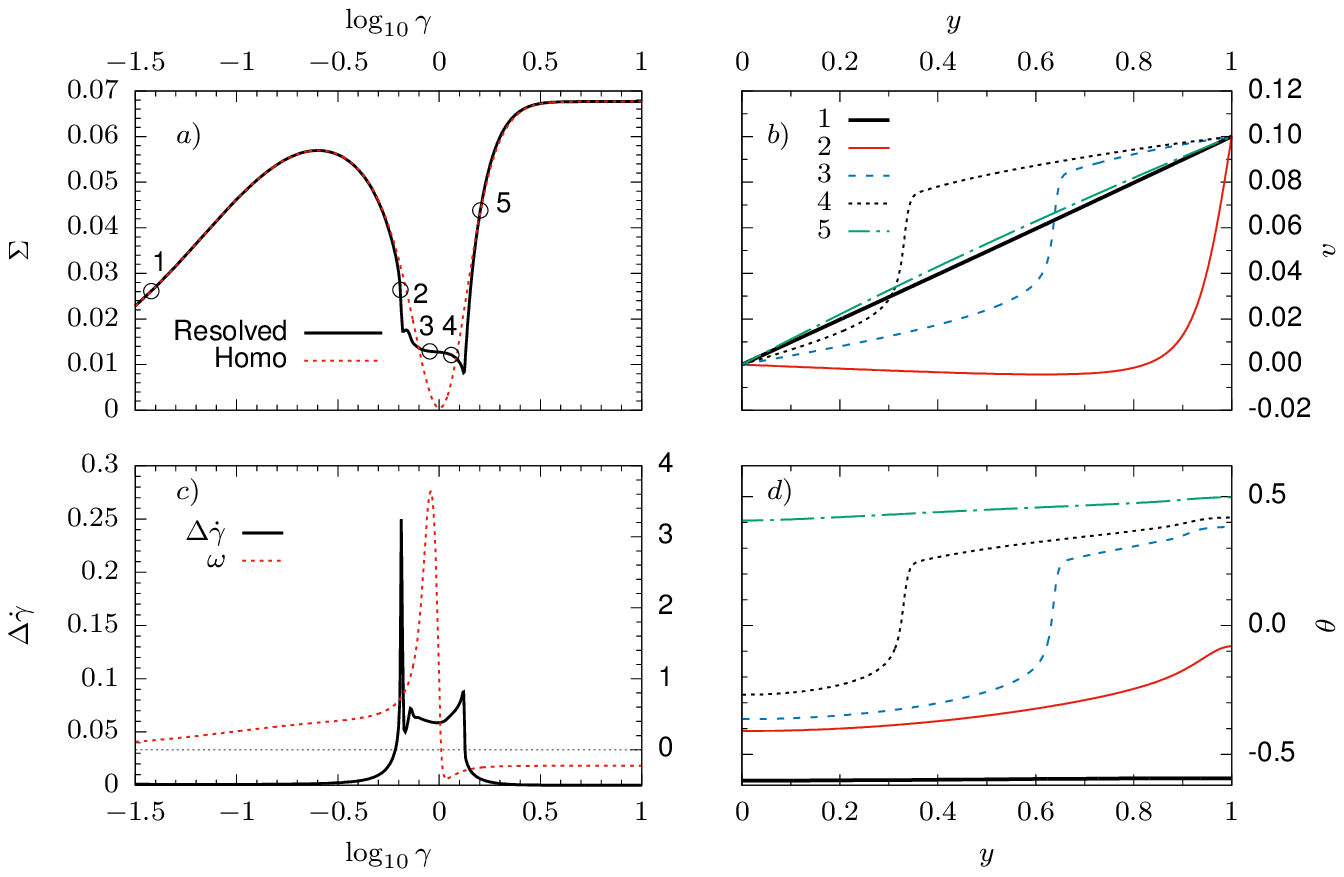}
\end{center}
\caption{Spatially resolved model for imposed average shear rate
  $\overline{\dot{\gamma}}=0.1$, polymer anisotropy $r=2$, and initial
  director angle $\theta_0 = -0.6$. a) Shear stress as a function of
  shear strain, b) velocity profile as a function of position $y$ at
  the time points labelled $1-5$ in a). The maximum real part of the eigenvalues,
  $\omega$, as a function of time (right hand $y$-axis), and shear rate
  inhomogeneity $\Delta \dot{\gamma}$ (left hand $y$-axis) are shown
  in c). The director angle is shown in d) for the corresponding lines
  shown in velocity profile plot b). }
\label{fig:spatialcomparison1}
\end{figure*}
The typical results of the calculation for imposed average shear rate
are shown in Fig.~\ref{fig:spatialcomparison1} for
$\overline{\dot{\gamma}}=0.1$, for an initial director angle of
$\theta_0 = -0.6$, i.e. with the director tilted away from the flow
direction. The shear stress in the spatially resolved model in
Fig.~\ref{fig:spatialcomparison1} a) follows the homogeneous
calculation initially. Once the director rotation starts then there is
a sharp dip in the shear stress, where the spatially resolved model
and the homogeneous model start to differ. The spatial shear rate then
becomes inhomogeneous as shown by $\Delta \dot{\gamma}$ in
Fig.~\ref{fig:spatialcomparison1} c). This coincides with the maximum
eigenvalue of the stability matrix, $\omega$. The velocity profile is
shown in Fig.~\ref{fig:spatialcomparison1} b) for various shear strain
values indicated in a). They show a high strain rate band propagating
across the rheometer gap. The high shear rate region corresponds to
the rotation of the director as can be seen from
Fig.~\ref{fig:spatialcomparison1} c).

\begin{figure}[!htb]
\begin{center}
  \includegraphics[width =0.45\textwidth]{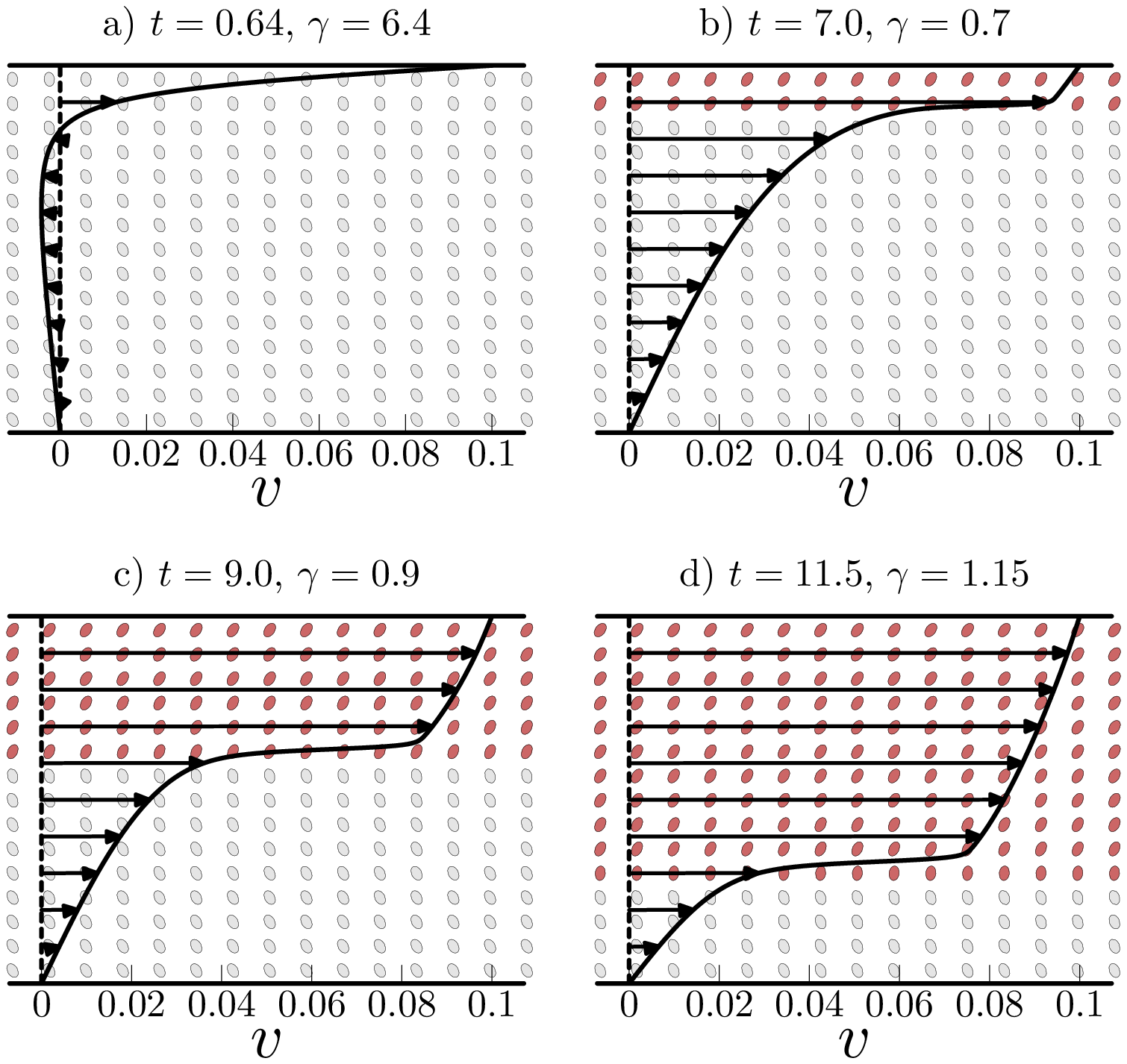}
\end{center}
\caption{An illustration of the velocity profiles for imposed average
  shear rate $\bar{\dot{\gamma}} = 0.1$, and initial director
  orientation $\theta_0 = -0.6$. The velocity field and the
  orientation of the director are shown as a function of space various
  for different time points in (a-d). The regions with the director
  pointing in the flow direction are shown with dark (red) shaded
  ellipsoids, and those with the director oriented away from the flow
  direction are shown in light grey.}
\label{fig:spatialgdotquadrics01}
\end{figure}

This picture is shown more clearly in
Fig.~\ref{fig:spatialgdotquadrics01}. Here the polymer conformation
tensor $\ten{W}$ is represented by an ellipsoid. This illustrates the
director orientation, and the local anisotropy. At the onset of
rotation shown in a) almost the whole fluid becomes stationary, and a
high strain rate region develops next to the wall. This high strain
rate region propagates across the rheometer rotating the
director. After the director has rotated the local strain rate drops
dramatically, resulting in plug flow.

\begin{figure}[!htb]
\begin{center}
\includegraphics[width = 0.45\textwidth]{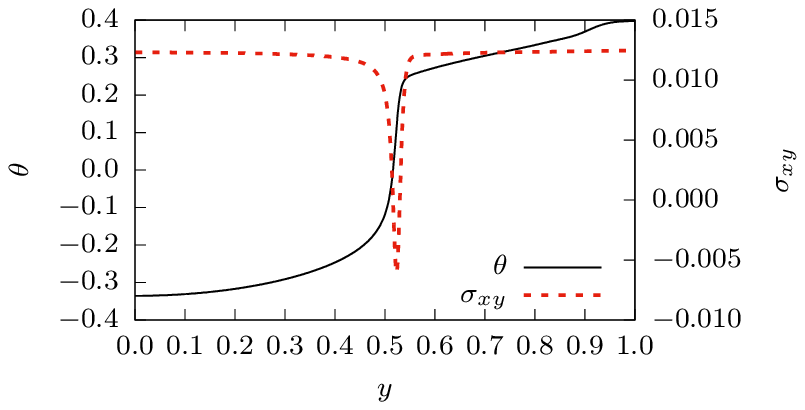}
\end{center}
\caption{The spatial dependence of the director angle, the polymer
  stress $\sigma_{xy}$ for average shear rate
  $\overline{\dot{\gamma}} = 0.1$, with initial condition
  $\theta_0 = -0.6$, at the time point $t= 1.0$, $\gamma = 10.0$. Note
  that there is a sharp drop in the polymer stress where the director
  rotation occurs.}
\label{fig:fixed_shearrate_stress_angle}
\end{figure}

\begin{figure}[!htb]
\begin{center}
\includegraphics[width = 0.45\textwidth]{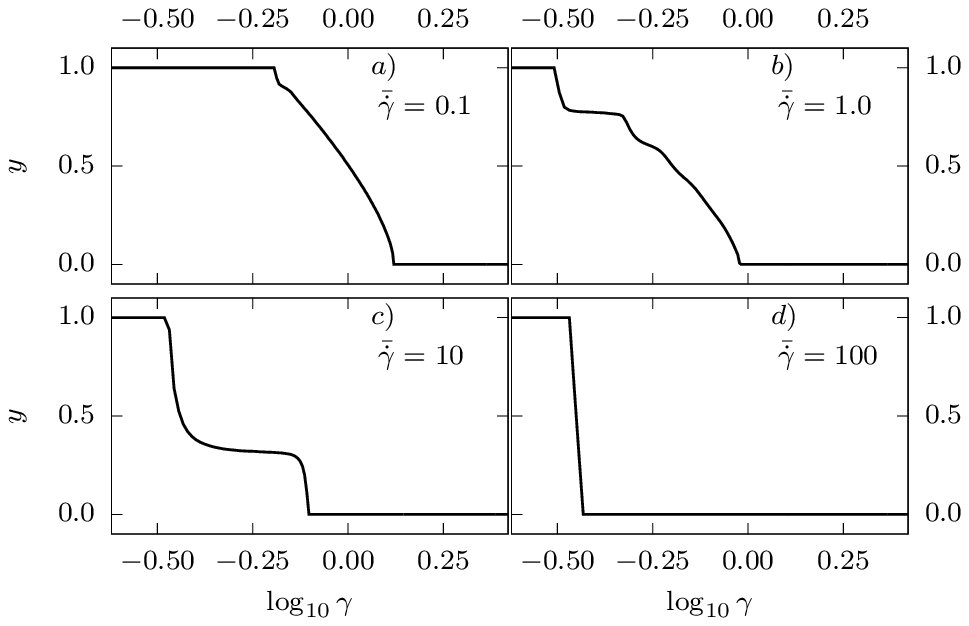}
\end{center}
\caption{The position in the gap $y$ of the boundary between the region
  where the director points in the flow direction ($0<\theta<\pi/2$)
  as a function of strain $\gamma$ for initial director orientation
  $\theta_0=-0.6$, for different values of imposed average shear rate
  $\overline{\dot{\gamma}}$ shown on plots (a-d). Note that for high
  average shear rates the rotation of the director is almost
  simultaneous for all $y$. }
\label{fig:vel_profiles}
\end{figure}

The mechanics of the director rotation can be seen clearly by plotting
the director angle and the shear stress on the same axes, as shown in
Fig.~\ref{fig:fixed_shearrate_stress_angle}. The polymer component of
shear stress $\sigma_{xy}$ drops dramatically at spatial point where
the director is rotating. This drop in stress during director rotation
is typical of liquid crystalline polymer systems. The total stress
across the sample is fixed, so there is a corresponding rise in the
shear rate, and hence the viscous component of the shear stress. The
highly sheared region propagates across the gap causing director
rotation.
The director rotation is particularly pronounced when
$\dot{\gamma}\sim 1$. For much higher shear rates the rotation front
propagates very rapidly across the sample, and director rotation
occurs simultaneously for all values of $y$. This is the elastic limit
of the ND model. A range of flow behaviour is shown in
Fig.~\ref{fig:vel_profiles} where the boundary between the rotated and
the unrotated director regions is illustrated.

\begin{figure}[!htb]
\begin{center}
  \includegraphics[width =0.45\textwidth]{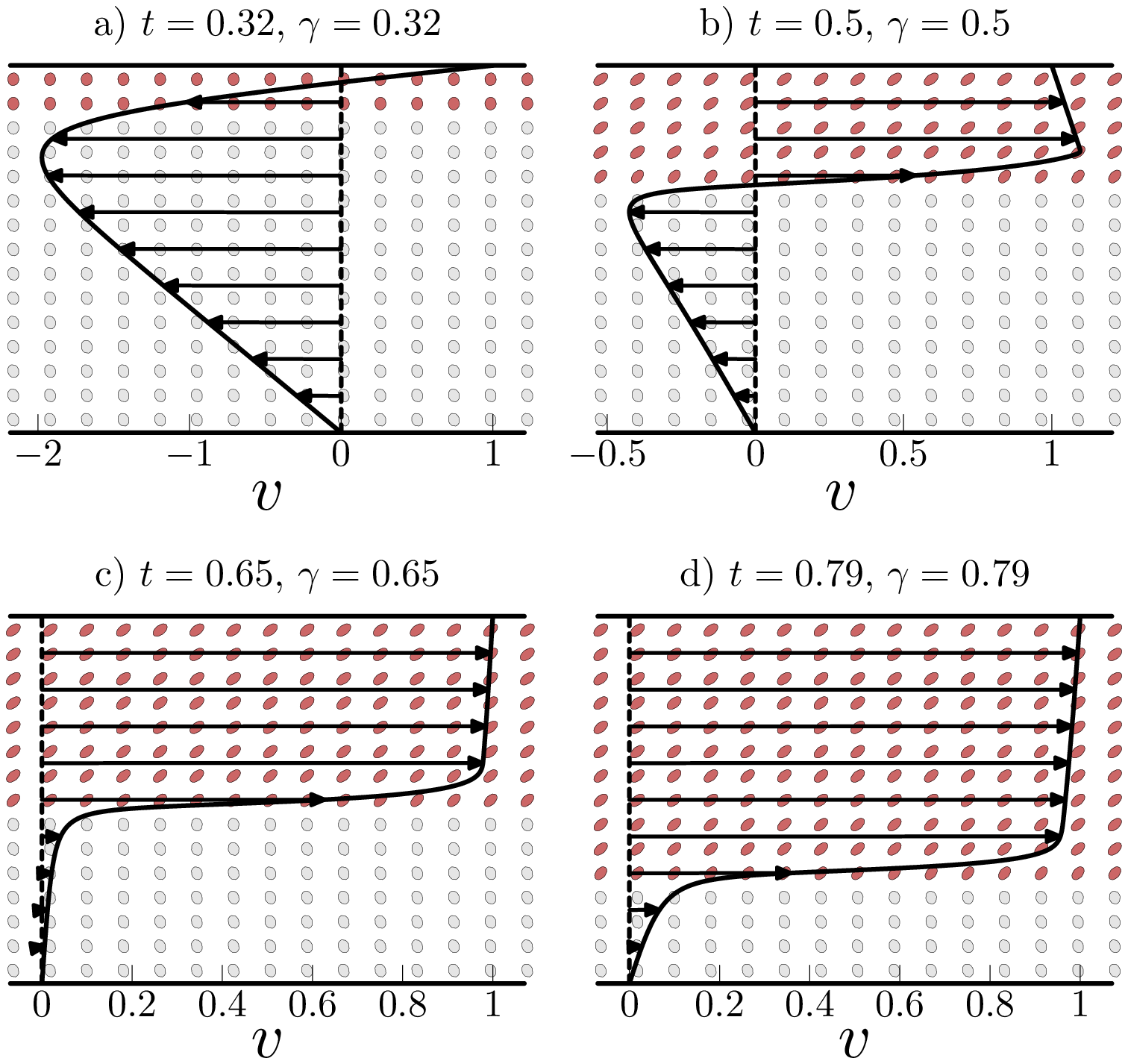}
\end{center}
\caption{An illustration of the velocity profiles, and polymer shape
  tensor for $\overline{\dot{\gamma}}=1$ and $\theta_0 = -0.6$. The
  time of each velocity profile is shown above each plot (a-d). The
  regions with the director pointing in the flow direction are shown
  with dark (red) shaded ellipsoids, and those with the director
  oriented away from the flow direction are shown in light grey. }
\label{fig:spatialquadrics1}
\end{figure}

For higher shear rates the flow profile can show recoil
behaviour. This is shown in Fig.~\ref{fig:spatialquadrics1} for
$\overline{\dot{\gamma}}=1$. At the onset of director rotation the
drop in the shear stress from rotation requires a negative velocity in
the rest of the sample to produce the required shear rate. The
interface between the rotated and the unrotated phases is much more
sharply defined here, resulting in plug flow -- i.e. the whole rotated
phase moves with the same velocity.

\subsection{Imposed shear stress}

\begin{figure*}[!htb]
\begin{center}
  \includegraphics[width=\textwidth]{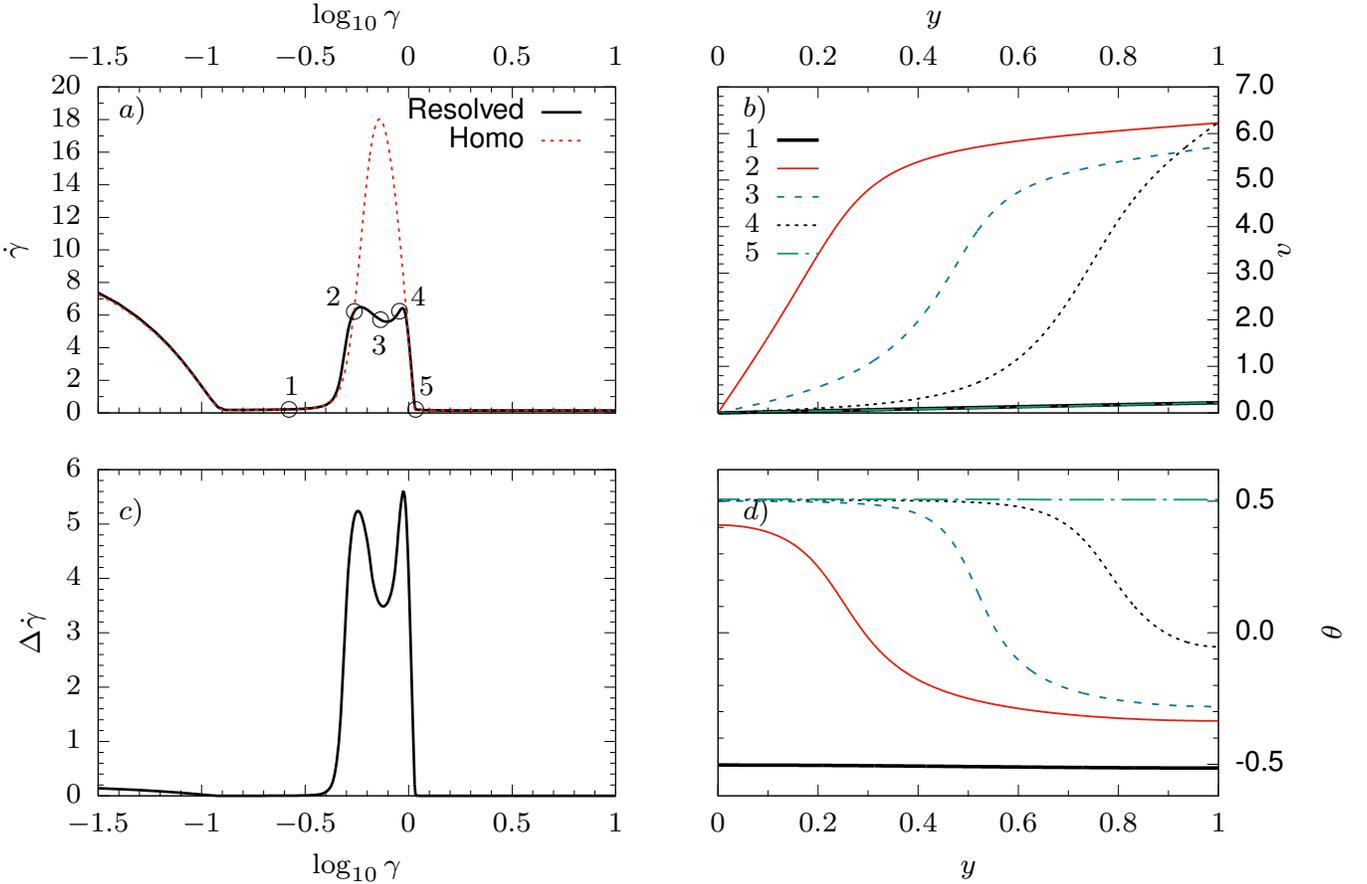}
\end{center}
\caption{Spatially resolved calculations for imposed total shear
  stress $\Sigma=0.1$, initial director angle $\theta_0 = -0.6$, and
  polymer anisotropy $r=2$. a) shows the average shear rate as a
  function of average shear strain for the spatially resolved and
  homogeneous calculations. b) shows the velocity profiles as a
  function of position across the gap, $y$, for the strain values
  indicated in a). c) shows the measure of inhomogeneity in shear rate
  $\Delta \dot{\gamma}$ as a function of average shear strain. d)
  shows the director angle as a function of spatial position $y$ for
  the corresponding strain values indicated in b). }
\label{fig:spatial_sigma}
\end{figure*}

Typical results of the spatially resolved calculation for fixed
imposed shear are shown in Fig.~\ref{fig:spatial_sigma}. Here figure \ref{fig:spatial_sigma} a)
shows the average shear rate for the spatially resolved and the
spatially homogeneous calculations. They are identical for small
strains. The degree of spatial inhomogeneity can be seen in
\ref{fig:spatial_sigma} b). Once the velocity profile becomes
inhomogeneous then the shear rates in a) differ -- the spatially
resolved system has a much lower average shear rate. The corresponding
spatial profiles for the velocity and director angle are shown in b)
and d) respectively. This shows that a high shear rate front
propagates across the rheometer gap, accompanied by a rotation of the
director. Once the director has rotated to the steady state value,
then the average shear rate drops sharply, and is consistent with the
spatially homogeneous results.  

\begin{figure}[!htb]
\begin{center}
  \includegraphics[width =0.45\textwidth]{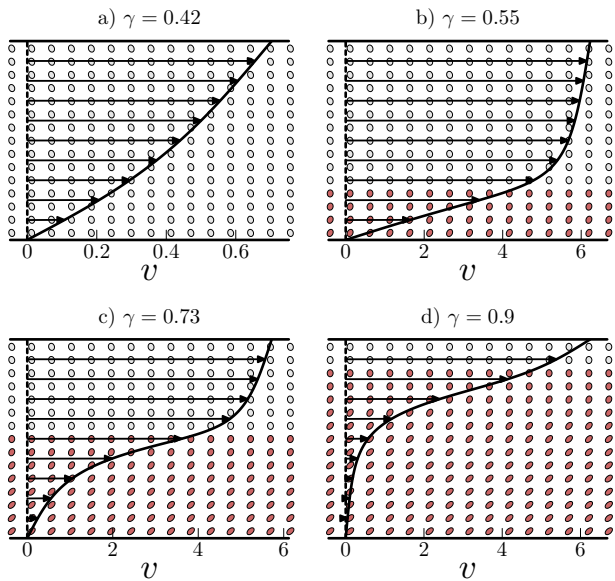}
\end{center}
\caption{The velocity profiles at different values of average strain,
  $\gamma$ for an imposed total shear stress $\Sigma = 0.1$, and
  initial director orientation $\theta_0 = -0.6$. The ellipsoids
  indicate the conformation of the polymer. The angle of the principal
  axis corresponds to the director orientation. The regions with the
  director pointing in the flow direction are shown with dark (red)
  shaded ellipsoids, and those with the director oriented away from
  the flow direction are shown in light grey.}
\label{fig:spatialsigmaquadrics}
\end{figure}

The velocity distribution and the polymer conformation are also shown
in Fig.~\ref{fig:spatialsigmaquadrics} for a range of different shear
strains. Here it can be seen that the rotation front nucleates at the
stationary plate of the rheometer ($y=0$) in b). This front is
associated with a high shear rate that flips the orientation of the
director. Once the director is rotated then it has a much lower
velocity.

\subsection{Flow reversal}

\begin{figure}[!htb]
\begin{center}
  \includegraphics[width =0.5\textwidth]{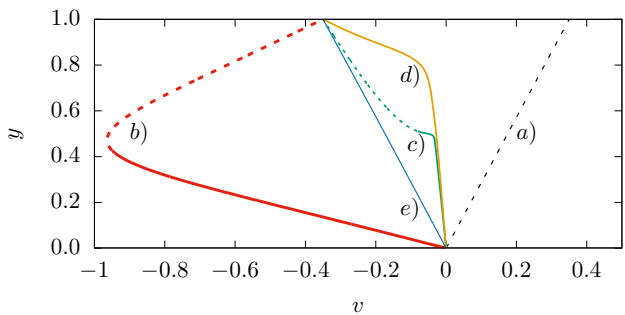}
\end{center}
\caption{The velocity profiles for flow-reversal protocol with
  $\overline{\dot{\gamma}}=0.35$ for $t\leq 14$ and then
  $\overline{\dot{\gamma}}=-0.35$ for $t>14.0$. The lines labelled
  $a)-e)$ correspond to $t = 13.0, 16.1, 16.4, 16.8, 16.9, $ and
  $20.0$ respectively. A solid line indicates that the director angle
  $\theta>\pi/2$ and a dashed line indicates that $\theta<\pi/2$.}
\label{fig:flowrev}
\end{figure}

The flow instability here in start up from rest depends critically on
the initial condition. This is not practical for experimental
systems. However, flow-reversal experiments are more practical to
carry out in LCPs and have observed a change in the order parameter on flow
reversal \cite{ugaz2001, mather2000}. To illustrate the behaviour of
the ND model under flow reversal the initial conditions were set with
the director close to its steady state value: $\theta_0 =0.6$. A fixed
average shear rate of $\overline{\dot{\gamma}} = 0.35$ was then
applied from $t=0$ to $t=14$, at which point it was reversed to
$\overline{\dot{\gamma}} = -0.35$. The results of the calculation are
shown in Fig.~\ref{fig:flowrev}. The resulting inhomogeneous velocity
profile is very similar to that observed in start up shear -- an
inhomogeneous shear rate develops, then a high shear rate front
propagates across the gap coinciding with director rotation. This may
be a more practical experimental test for this theory.

\section{Discussion}
\label{sec:disc}

The ND constitutive model is a logical extension of the upper
convected Maxwell (UCM) model, and describes semi-flexible LCPs, i.e.
where each polymer chain can be distorted by the flow field. The
calculations presented here show that this model has a transient flow
instability to the formation of an inhomogeneous velocity profile
under certain initial conditions. Its behaviour is qualitatively
different to the shear banding observed in models that describe
worm-like micellar solutions and polymer solutions, such as the
Diffusive Johnson-Segalmann (DJS) model \cite{doi:10.1122/1.551085},
and the Vasquez-Cook-McKinley (VCM) model
\cite{doi:10.1122/1.2829769}. These models are constructed to have
shear banding in the steady state through a non-monotonic constitutive
curve. The flow forms two bands -- a high shear rate aligned phase and
a low shear rate isotropic phase -- with the average shear rate
imposed on the system. The transient velocity profiles in models of
polymer solutions such as the Diffusive Rolie-Poly (DRP) model
\cite{PhysRevLett.102.067801} does not require a non-monotonic
constitutive curve, but still has the same form of a high shear rate
and a low shear rate band.

The ND model has a monotonic constitutive curve, but exhibits a
different type of inhomogeneous velocity profile to transient
shearbanding in the DRP model. A high shear rate front propagates
across the rheometer gap and induces director rotation. This model is
dominated by the elasticity of the polymer chains, hence the defect
dynamics have no effect on the director distribution as observed in
models of rod-like LCPs \cite{PhysRevE.68.061704}. The ND model may
exhibit even richer behaviour in higher dimensions, such as banding in
the vorticity direction as well as the gradient direction, as has been
found in the DJS model \cite{PhysRevLett.96.104502}.

There is both experimental evidence of mechanically induced phase
transition in LCPs \cite{doi:10.1021/ma970737h}, and consistent
theoretical calculations \cite{PhysRevA.46.4966,PhysRevA.41.4578}. For
semi-flexible LCPs the calculations here suggest that measurement of
the order parameter should be done in such a way as to avoid averaging
over the spatial variation in the director induced by the flow. This
could arise if the measurements are taken by averaging across the
gradient direction in the rheometer, for example by X-ray scattering
with the beam passing through a Couette rheometer along the radial
direction. A possible experimental test for this model is to use
particle tracking velocimetry to measure the velocity distribution
during start up flow, or a flow-reversal experiment. This experiment
would reveal the inhomogenenous velocity profile predicted by the ND
model.

The dynamics of the director rotation in this model are closely
related to the formation of microstructure in liquid crystal
elastomers \cite{MACP:MACP677}. Here the typical geometry is an
elongational deformation. Stripe domains of alternating rotation in
the director field form. Imposed elongational flow in the ND model
might produce microstructure with similar striped domains in the
velocity profile.

Using mixtures of oblate and prolate chains could be modelled using
the ND model to create LCPs with a tuneable flow aligning behaviour
\cite{PhysRevLett.90.115501}.

\section{Conclusion}

We have analysed the nematic dumbbell model of Marrucci and Maffetone
\cite{Maffetone1992} with an additional polymer diffusion term, and a
Newtonian solvent term. By using a linear stability analysis we
determined the effect of spatial perturbations in the polymer stress
components. These calculations were performed for both fixed shear
strain rate, and fixed total shear stress. For initial conditions
where the director is rotated away from the flow direction linear
stability analysis shows that it is unstable. Spatially resolved
calculations of the velocity profile show that there is some spatial
structure in the velocity profile which corresponds to the
reorientation of the director during the flow. The director rotation
is confined to a front that propagates across the gap in the
rheometer. For high imposed shear strain rates, or high total shear
stress the rotation of the director occurs almost simultaneously
across the whole sample. These calculations suggest that investigation
of the spatial structure of the velocity field in the rheology of
semi-flexible flow aligning liquid crystalline polymers may yield
interesting results. One possible experimental test of this prediction
is to use particle tracking velocimetry to measure the velocity
profile of semi-flexible liquid crystalline polymers across the gap of
a couette rheometer during a start up shear experiment.


\appendix
\section{Elastic Limit}
\label{app:elasticlimit}

\begin{figure}
\begin{center}
\includegraphics[width = 0.5\textwidth]{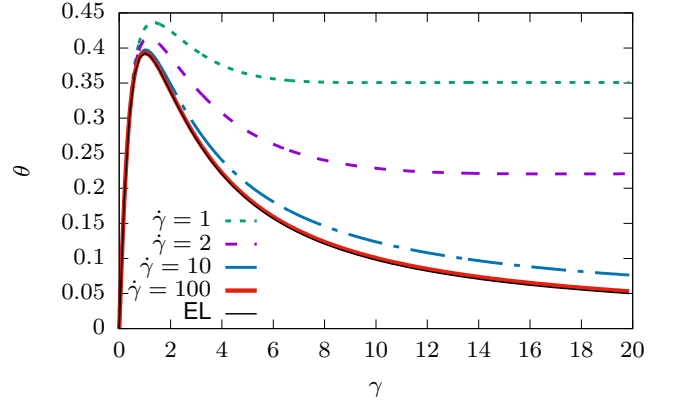}
\end{center}
\caption{Reorientation of a dumbbell initially aligned along the flow
  direction in response to a shear strain $\gamma$ for various values
  of $\dot{\gamma}\tau_{\perp}$. The elastic limit (shown by the solid
  black line) corresponds to
  $\dot{\gamma}\tau_{\perp}\rightarrow\infty$.}
\label{fig:ellimit}
\end{figure}

In the limit $t<<\tau_{\perp}$ the response of the system to an
imposed shear strain should be purely elastic.  We can thus ignore the
viscous terms in Eq.~(\ref{eqn:totalstress}). The constitutive
equations are then:
\begin{eqnarray}
\dot{W}_{xx}&=&2W_{xy}\dot{\gamma}\\
\dot{W}_{yy}&=&0\\
\dot{W}_{xy}&=&W_{yy}\dot{\gamma}.
\end{eqnarray}
Integrating these equations for a constant shear strain rate we
obtain:
\begin{eqnarray}
W_{xx}(t)&=&W_{xx}(0)+2\gamma(t) W_{xy}(0)+\gamma(t)^{2}W_{yy}(0)\\
W_{xy}(t)&=&W_{xy}(0)+\gamma(t)W_{yy}(0)\\
W_{yy}(t)&=&W_{yy}(0)
\end{eqnarray}
where the strain is given by $\gamma(t)=\dot{\gamma}t$. The director
at a strain $\gamma$ is denoted by $\vec{n}=(\cos\theta,\sin\theta)$
and is the eigenvector associated with the largest eigenvalue of
$\ten{W}$, it is simple to show that $\theta$ satisfies:
\begin{equation}
\tan 2\theta(t)=\frac{2W_{xy}(t)}{W_{xx}(t)-W_{yy}(t)}.\label{eq:angles}
\end{equation}
For the initial values we assume a nematic with anisotropy $r$ and
initial director aligned along
$\vec{n}_{0}=(\cos\theta_{0},\sin\theta_{0})$, thus:
\begin{eqnarray}
W_{xx}(0)&=&1+(r-1)\cos^{2}\theta_{0}\\
W_{xy}(0)&=&(r-1)\sin\theta_{0}\cos\theta_{0}\\
W_{yy}(0)&=&1+(r-1)\sin^{2}\theta_{0}
\end{eqnarray}
using these values and solutions for $W_{xx}(t)$, $W_{xy}(t)$ and
$W_{yy}(t)$ above we obtain the dependence of the angle $\theta$ on
the shear strain $\gamma$ in the elastic limit
\begin{equation}
\tan 2\theta(\gamma)=\frac{\sin(2\theta_{0})-\gamma\cos(2\theta_{0})+\gamma\frac{(r+1)}{(r-1)}}{(1-\gamma^{2}/2)\cos(2\theta_{0})+\gamma\sin(2\theta_{0})+\frac{(r+1)}{(r-1)}\frac{\gamma^{2}}{2}}
\end{equation}
This limit should describe the reorientation of the director for
strains less than $\gamma\sim \dot{\gamma}\tau_{\perp}$.  In
Fig.~\ref{fig:ellimit} we plot the reorientation of a nematic with
$\theta_{0}=0$ for various values of $\dot{\gamma}$ as a function of
strain $\gamma$. As can be seen for small strains the reorientation
follows the elastic limit (black line), but for strains
$\gamma \gtrsim \dot{\gamma}\tau_{\perp}$ we start to see deviations
from the elastic limit as stress begins to relax viscously.

\section{Small strain response}
\label{app:saresponse}

We work here in two dimensions, writing the director and its
perpendicular component as
\begin{eqnarray}
\vec{n} &=& ( \cos \theta, \sin \theta)\\
\vec{n}_\perp &=& (-\sin \theta, \cos \theta)\\
\Rightarrow \ten{W} &=& (r+\delta) \vec{n} \vec{n} + (1+\epsilon) \vec{n}_\perp \vec{n}_\perp
\end{eqnarray}
where $\delta$ and $\epsilon$ are the leading order changes in the
diagonal components for small amplitude shear. We will apply a
velocity gradient given by $\ten{K} = \dot{\gamma}\hat{\vec{x}}
\hat{\vec{y}}$.
The components of the constitutive equations can then be calculated by
taking the appropriate dot products
$\vec{n}\cdot \ten{W}\cdot\vec{n}$,
$\vec{n} \cdot\ten{W} \cdot\vec{n}_\perp$ and
$\vec{n}_\perp \cdot\ten{W} \cdot\vec{n}_\perp$. Using this basis
results in the following equation for the polymer stress
$\sigmabold = \ellbold^{-1} \cdot \ten{W}$,
\begin{equation}
\sigma_{xy} = \frac{\delta - r \epsilon}{2r}\sin 2 \theta
\end{equation}
and the following equations result from the components of the
constitutive equation.
\begin{eqnarray}
\dot{\delta}&=&- \frac{2 \delta}{r \tau}+ \dot{\gamma}(r+\delta) \sin 2 \theta\\
\dot{\epsilon}&=& - \frac{2 \epsilon}{\tau}- \dot{\gamma}(1+\epsilon)\sin 2 \theta\\
\dot{\theta}&=&\frac{1}{2}\dot{\gamma}\frac{1-r-\delta+\epsilon+(1+r+\delta+\epsilon)\cos 2 \theta}{r+\delta - 1 - \epsilon}
\end{eqnarray}
These can be solved to find the leading order response for small
deviations of $\theta$ from its starting orientation
\mbox{$\theta =\theta_0+ \xi$} under oscillatory shear strain
$\gamma(t) = \gamma_0 \sin \omega t$. In this case
\begin{eqnarray}
\dot{\xi} &\approx& \frac{\gamma_0(1-r+(1+r) \cos 2 \theta_0)}{2 (r-1)} \omega\cos \omega t\\
&-&\gamma_0 \delta\omega \cos \omega t\frac{\cos 2 \theta_0}{(r-1)^2}+\gamma_0\epsilon\omega \cos \omega t\frac{r\cos 2 \theta_0}{(r-1)^2}\\
&-&\xi\gamma_0\omega \cos \omega t\frac{(r+1)\sin 2 \theta_0}{r-1}.
\end{eqnarray}
Note that when $\cos 2\theta_0 = \frac{r-1}{r+1}$ then the leading
order in $\xi$ is zero.

Leading order response is
\begin{equation}
\xi(t) =  \frac{\gamma_0(1-r+(1+r) \cos 2 \theta_0)}{2 (r-1)\omega} \omega \sin \omega t
\end{equation}
Substituting this back into the equations for $\delta$ and $\epsilon$,
we find the leading order response for the shear stress in the limit
$t\rightarrow \infty$ (after the transient has dissipated).
\begin{eqnarray}
\sigma_{xy} &=& \frac{\gamma_0 \omega\sin^2 2 \theta_0}{(4+ \omega^2)(4+r^2 \omega^2 )}\times\nonumber\\
&&  \left( \omega (2 + r^2(2+ \omega^2))\sin  \omega t\right.\nonumber\\
&+& \left.(1+r)(4+r \omega^2 )\cos\omega t\right).
\label{eqn:smallstrain}
\end{eqnarray}
We can extract the storage and loss modulus from
Eq.~(\ref{eqn:smallstrain}): 
\begin{eqnarray}
G^\prime(\omega) &=& \frac{\omega^2  (2+r^2(2 +  \omega^2))}{(4 +\omega^2)(4 + r^2 \omega^2)}\sin^2 2 \theta_0\\
G^{\prime \prime}(\omega) &=&\frac{\omega   (r+1)}{4 +  \omega^2}\sin^2 2 \theta_0.
\end{eqnarray}
Note this material becomes soft (i.e. $G' = G'' = 0$) when $\theta_0$
is small, but the analysis is not valid for $\theta_0=0, \pi/2$. This
is the soft elastic response observed in LCEs as a result of the
rotation of the director \cite{warner2003}. The isotropic results
$r=1$ of the Upper Convected Maxwell model can be recovered by setting
$\theta_0=\pi/4$ and $r=1$.

When $\theta_0=0$ then the response becomes much softer and is no
longer sinusoidal.
\begin{eqnarray}
\sigma_{xy} &=& \frac{\gamma_0^3 \omega \cos \omega t}{(r-1)^2(1+\omega^2)(1+r^2  \omega^2)}\times\nonumber\\
&&\left( \omega (1+r^2(1+2  \omega^2))\sin 2 \omega t \right.\nonumber \\
&&\left. + (1+r) (1+r \omega^2)\cos 2 \omega t\right)
\end{eqnarray}
The material has no linear response regime here due to the soft
rotation of the director. This contains both the $\omega t$ and
$3 \omega t$ harmonics at the same order in $\gamma_0$. This
degeneracy in the model could be removed by including the response of
the Newtonian solvent term, or by modifying the constitutive equation
of the LCP to include imperfections such as the dispersity of the
anisotropy as has been done for semi-soft LCEs.

For larger amplitude oscillatory shear the response is non-linear due
to the rotation of the director during the flow.

\section{Integration of fluctuations}
\label{app:gdotfluct}

\begin{figure}[!htb]
\begin{center}
  \includegraphics[width =
  0.48\textwidth]{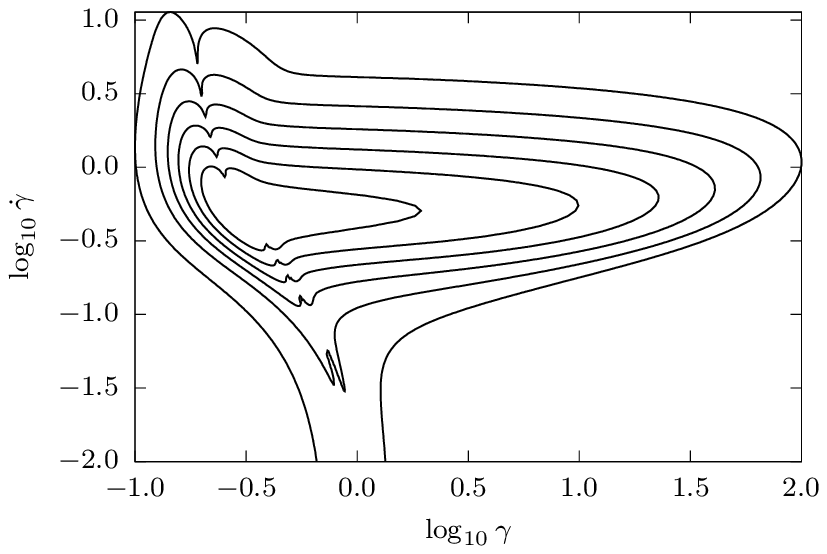}
\end{center}
\caption{The integrated fluctuation in the shear rate
  $\delta \dot{\gamma}$ from Eq.~(\ref{eqn:dgamfluctuations}) for
  $\theta_0 = -0.6$. The lines correspond to
  $\log_{10} \delta \dot{\gamma}= -1, 1,3,5,7,9$, for an initial
  amplitude of $\delta W_{ij} = 10^{-3}$.}
\label{fig:dgdotintegrated}
\end{figure}

The extent of the growth in fluctuations during the shear flow can be
measured using the shear rate fluctuations from
Eq.~(\ref{eqn:dgamfluctuations}). We can integrate the result over
time (or equivalently strain). This approach was followed in
Ref.~\cite{:/content/sor/journal/jor2/58/1/10.1122/1.4842155}. The
solution of the constitutive equations was first calculated in the
eigenbasis. The LSA was done in the \emph{Cartesian} basis and the
fluctuations in $\dot{\gamma}$ integrated using the initial conditions
of $\delta W_{xx}, \delta W_{xy}$ and $\delta W_{yy}$ set to
$10^{-3}$. The NAG C library routine \texttt{d02ejc} was used to
integrate these equations.  Fig.~\ref{fig:dgdotintegrated} shows the
result of this calculation for the ND model, for an unstable initial
configuration of $\theta_0 = -0.6$. As can be seen from the contours
in this figure the fluctuations grow most strongly for
$\dot{\gamma}\sim 1$. The cusp running down the contours arises from
the change in sign of $\delta{\dot{\gamma}}$ during the
calculation. The fluctuations eventually decay away indicating that
the instability in this model is transient, and the steady state is
spatially homogeneous.

\section{Properties of LSA eigenvalues}
\label{app:LSAeigenvalues}

\begin{figure}
\begin{center}
\includegraphics[width = 0.5\textwidth]{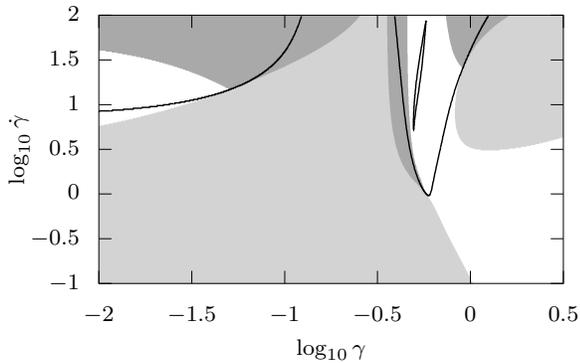}
\end{center}
\caption{This figure shows the number of eigenvalues of the stability
  matrix $\ten{P}$ in Eq.~(\ref{eqn:stabmat}) with positive real part
  for imposed average shear rate $\dot{\gamma} = 0.1$ and
  $\theta = -0.6$.  White corresponds to $0$ eigenvalues with positive
  real part, light grey to $1$ and dark grey to $2$. The black lines
  enclose the region where there is a Hopf bifurcation, i.e. two
  eigenvalues are complex conjugates pairs. }
\label{fig:hopf}
\end{figure}

A general criterion for the determination of the stability of the
flow, for the fixed shear rate case, based on LSA has been derived in
\cite{:/content/sor/journal/jor2/58/1/10.1122/1.4842155}:
\begin{equation}
  \epsilon- G \vec{p}\cdot \vec{M}^{-1}\cdot \vec{q}<0.
\end{equation}
Some of the assumptions used in developing this criterion are not
satisfied by the ND model. Firstly it is assumed that the determinant
of $\ten{M}$ in Eq.~(\ref{eqn:dsfluctuations}) obeys
$(-1)^D |\ten{M}|<0$, where $D$ is the dimensionality of
$\ten{M}$. Whilst it can be shown that the determinant is negative in
equilibrium, it does change sign as the ND model evolves, and depends
on the applied shear rate. The eigenvalues of $\ten{M}$ are all real
for small values of $\dot{\gamma} \approx 0.1$. For larger values of
shear rate there is a Hopf bifurcation, and corresponding complex
eigenvalues. In this case the determinant changes sign from negative
to positive, and then back to negative. This behaviour of the
eigenvalues means that analysing the determinant of $\ten{M}$
(i.e. the product of the eigenvalues) is not enough to determine if
one of them has changed sign. The real part of two of the three
eigenvalues could change sign simultaneously (in the Hopf
bifurcation), and leave the sign of the determinant unchanged.
Secondly the determinant of $\ten{P}$ of Eq.~(\ref{eqn:stabmat}) also
shows a Hopf bifurcation. Fig.~\ref{fig:hopf} shows the eigenvalues of
$\ten{P}$. The shading here shows that there are regions of $0,1$ or
$2$ eigenvalues that have positive real part respectively. Some of the
regions with $0$ or $2$ eigenvalues of positive real part can have
complex conjugate pairs of eigenvalues -- a Hopf bifurcation. These
regions are indicated by the black line.

\begin{figure}[!ht]
\begin{center}
\includegraphics[width = 0.5\textwidth]{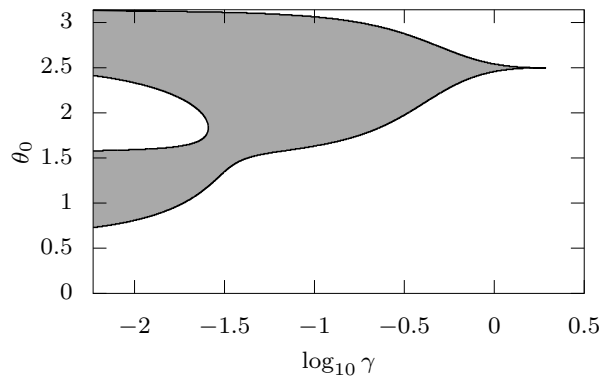}
\end{center}
\caption{The shaded region shows where the maximum eigenvalue of
  $\ten{P}$ has a positive real part. The change in stability at
  $\gamma=0$ can be found from the $\dot{\theta}$ equation. The cusp
  for large $\gamma$ is the unstable result from the steady state
  equation. }
\label{fig:fixedshearrate_angle}
\end{figure}

Fig.~\ref{fig:fixedshearrate_angle} shows the dependence of the
maximum real part of the eigenvalue on the starting angle,
$\theta_0$. The system is unstable for large strains in the region of
$\theta_0>\pi/2$. There is a cusp for large strain at an angle
corresponding to the unstable director orientation from the steady
state solution.

\end{document}